# Media Integrity and Authentication: Status, Directions, and Futures

Jessica Young[†], Sam Vaughan, Andrew Jenks, Henrique Malvar, Christian Paquin, Paul England, Thomas Roca, Juan LaVista Ferres, Forough Poursabzi, Neil Coles, Ken Archer, Eric Horvitz[†]

Microsoft

**Abstract**
We provide background on emerging challenges and future directions with *media integrity and authentication* methods, focusing on distinguishing AI-generated media from authentic content captured by cameras and microphones. We evaluate several approaches, including provenance, watermarking, and fingerprinting. After defining each method, we analyze three representative technologies: *cryptographically secured provenance*, *imperceptible watermarking*, and *soft-hash fingerprinting*. We analyze how these tools operate across modalities and evaluate relevant threat models, attack categories, and real-world workflows spanning capture, editing, distribution, and verification. We consider sociotechnical "reversal" attacks that can invert integrity signals, making authentic content appear synthetic and vice versa, highlighting the value of verification systems that are resilient to both technical and psychosocial manipulation. Finally, we outline techniques for delivering *high-confidence provenance authentication*, including directions for strengthening edge-device security using secure enclaves.

**About this report**
This report draws on findings from a study conducted over several months in 2025 under Microsoft's *Longer-term AI Safety in Engineering and Research* (LASER) program. Led by the Office of the Chief Scientific Officer (OCSO), LASER studies convene experts from across Microsoft to examine potential concerns and challenges arising from rapidly evolving AI technologies, including foundational advances and new categories of AI applications that introduce novel capabilities, behaviors, or potential societal impacts.

[†]Contacts: {jessicayoung, horvitz}@microsoft.com

## Table of Contents

# Executive Summary

The *Media Integrity and Authentication: Status, Directions, and Futures* report is a study commissioned by the Office of the Chief Scientific Officer to explore the capabilities and limitations of *media integrity and authentication* (MIA) technologies—specifically provenance, watermarking, and fingerprinting. The study investigates vulnerabilities in these core technologies and identifies directions forward amidst technical and sociotechnical challenges. A primary goal was gaining transparency into how these methods perform under a range of realistic scenarios, with a focus on their resilience against adversarial attacks. Contributors to this study include a multidisciplinary team with expertise in AI, security, social sciences, human-computer interaction, policy, operations, and governance.

Microsoft is a pioneer in media provenance technology, built on early foundational research and its leadership in co-founding and advancing the Coalition for Content Provenance and Authenticity (C2PA). A Microsoft Research team, led by Chief Scientific Officer, Eric Horvitz, envisioned and prototyped an approach to certify authentic media, including news reports, videos, and transcripts, ensuring content from a trusted source remains untampered on its journey to the consumer. This work evolved into today's C2PA Content Credentials, which now includes expanded use cases such as disclosing AI-generated content.

With growing demand for authentication tools, increasing adoption of the C2PA standard, and emerging legislation requiring MIA technology, there is an important opportunity to more deeply understand how these tools can distinguish authentic content from increasingly sophisticated deepfakes. At the same time, we think it is critical to assess the limitations of these technologies to avoid overconfidence and overreliance.

We hope this report will serve as a resource for engineers, researchers, legislators, civil society organizations, and the public seeking to better understand the MIA ecosystem for images, audio, and video.[1] Specifically, the report addresses the following:

- What roles do provenance, watermarking, and fingerprinting play within the media integrity ecosystem?
- Where do these technologies succeed, and where do they fall short?
- How can they strengthen resilience to technical attacks?
- How can high-confidence results bolster resilience to sociotechnical attacks—deceptions that exploit how people understand, use, or trust these tools?
- Finally, what persistent "rough edges" should be anticipated?

In support of answering these questions, we find:

[1] This report focuses on image, audio, and video, which are currently more mature modalities for disclosing and validating AI-generated or modified content. Text remains an evolving area with unique challenges and complexities as discussed on page 8.

1. **Education Gap:** General confusion regarding the purpose and limitations of MIA methods highlights an urgent need for education. Expectations must be calibrated to the actual level of protection these technologies provide to appropriately inform policy and adoption.
2. **Regulatory Pressure:** While regulatory requirements are still being defined, legislation coming into effect in 2026 will require widespread use of media integrity methods. However, choices on implementation and display will directly impact the reliability of provenance indicators and how the public interprets them.
3. **Technical and Sociotechnical Vulnerabilities:** MIA methods can be susceptible to technical attacks as well as sociotechnical "reversal" attacks that are capable of inverting signals, making authentic content appear synthetic, and synthetic content appear authentic. Such attacks may mislead the public, resulting in widespread confusion about an asset's authenticity.
4. **Strength Through Layering:** Linking secure provenance with imperceptible watermarking enables *high-confidence validation,* the capability of verifying, under defined conditions, that claims about the origin of and modifications made to an asset can be validated with certainty. Recovering a C2PA provenance manifest created and signed in a high security environment with an imperceptible watermark ID offers a promising option to mitigate the impact of attacks and minimize confusion.
5. **Hardware-Level Security:** High-confidence results aren't feasible when provenance is added by a conventional offline device (e.g., camera or recording device without connectivity). To make the provenance of captured images, audio, and video trustworthy, it is essential to implement secure enclaves within the device hardware.
6. **Role of Fingerprinting:** Fingerprinting is not a viable path to high-confidence validation and faces significant scaling costs. However, it remains a valuable tool for *manual* forensics in high-risk scenarios requiring intensive assessment.
7. **Operational Utility:** All three methods have applications beyond online content. They offer organizations powerful tools for addressing operational challenges such as fraud prevention, risk management, and digital accountability.

**Strategic Directions**

To mitigate technical and sociotechnical attacks that could undermine trust in online content, and to inform critical implementation and policy decisions, the report focuses on four directions: (1) Deliver High-Confidence Authentication, (2) Mitigate Confusion from Sociotechnical Attacks, (3) Enable More Trusted Provenance on Edge Devices, and (4) Invest in Ongoing Research and Policy Development. A summary of considerations for each direction is provided below, and more detail is provided in each section of the report.

**Direction 1: Deliver High-Confidence Authentication**

- **Synthetic and Mixed Media** ▸ GenAI system providers should consider prioritizing provenance and watermarking for provenance recovery, where possible[2], for synthetic media generation and editing scenarios to enable high-confidence validation. To address cases involving heightened risk of abuse, organizations can explore provenance, watermarking, and fingerprinting to enable sequential authentication as needed.

- **Authentic Media** ▸ Organizations should recognize and explore uses of provenance for certifying and raising trust in authentic content and records (such as photos, transcripts, documents), including uses of provenance to capture history of changes made through editing and post-production.

- **Validation Tools** ▸ To minimize confusion and overreliance, we recommend provenance validation tool providers consider displaying only high-confidence results to the public. C2PA manifest validation and display should be the default way by which provenance information is shown on distribution platforms (e.g., social media sites) and publicly available first-party validation tools. Lower-confidence provenance results, if displayed, must be clearly distinguished from high-confidence indicators.

- **Accounting for Exceptions** ▸ As the use of secure provenance, for high-confidence results, won't be possible in all cases, industry should promote continued research and alignment on *display choices and media literacy*, to help mitigate legitimate, authentic media without provenance being discredited.

- **Forensic Access** ▸ Companies should consider making MIA services available for forensic investigators to access lower-confidence provenance signals that are not suitable for general public display.

- **Additional Safeguards** ▸ Due to security risks like potential "oracle attacks" on decoders, additional safeguards, such as employing multiple watermarks or unique keys, are necessary before making watermark detector tools publicly accessible.

[2] Maintaining flexibility will be necessary based on the scenario at hand. While prioritizing provenance supports high-confidence validation, there may be cases where provenance specifications (e.g., per the C2PA standard) have not been extended to account for use on new modalities. In other scenarios, watermarking may not be an effective solution. For instance, watermarking binary (black and white) images is also an evolving area, with a lack of robust techniques.

**Direction 2: Mitigate Confusion from Sociotechnical Attacks**

- **Region of Interest** ⏵ Verification site providers should consider displaying details about *where* edits occur within the media and, when possible, thumbnails of media inputs.
- **Manifest Preservation** ⏵ Distribution platforms (i.e., social media sites) should preserve details about where edits were made to media by enabling users to download complete manifest details or explore them via other tools.
- **UX Design** ⏵ The C2PA should push for research-based UX standards for consistent and effective provenance display across platforms. At the same time, regulators requiring perceptible markings should support the adoption of a standardized mark that is designed for consistent interpretation globally and to mitigate confusion when such marks are inevitably attacked.
- **Security** ⏵ C2PA must ensure that signing certificates accurately represent the security a hardware device or software application truly offers. Trusted signer lists that validation sites depend on must be updated regularly based on incident remediation.
- **State-of-the-Art (SOTA) Implementations** ⏵ Cameras should use secure metadata (e.g., secure implementations of C2PA-based provenance) to mitigate manipulated provenance information being displayed to content consumers. Online platforms consuming and relaying provenance information should, in turn, explore ways to differentiate between secure and insecure provenance information.

**Direction 3: Enable More Trusted Provenance on Edge Devices**

- **Disclosure in Low-Security Environments** ⏵ Device providers should explore using version 2.3 or a later version of the C2PA specification, which allows implementers to obtain signing certificates that reflect the security state of the environment for manifest generation and signing that occurs offline.
- **Display in Low-Security Environments** ⏵ Verification tools should show validated provenance information derived from offline devices for the highest confidence validation pathways (i.e., C2PA manifest validation, or watermark verification to recover a valid C2PA manifest). Validation tool providers should also explore displaying provenance information in a way that mitigates overreliance if the provenance was signed with a low security level.
- **Conformance and Display Alignment** ⏵ C2PA should carefully shape how security levels for provenance signing certificates impact provenance display.

  With the release of C2PA v2.3, the Conformance Program now defines and enforces security levels for provenance signing certificates. This enables platforms and validation tools to differentiate between high- and low-assurance provenance, and to next explore how to display this information to users in a consistent, research-informed manner. The Conformance Program's public registry of conformant products and Certification Authorities (CAs) supports ecosystem-wide interoperability and trust.

**Direction 4: Invest in Ongoing Research and Policy Development**

- **Use and Display Research** ▸ C2PA or its members should champion research workstreams to better understand the use and display of provenance signals both in the short- and long-term, and share these results with the community to improve consistency and effectiveness. Important research directions for display include in-stream tools that display provenance information where people are and distinguish between high- and lower-confidence provenance signals.
- **Manifest Stores** ▸ Further research is needed to define best practices for implementing manifest stores, including exploring a potential centralized collection of stores from multiple entities or a decentralized version.
- **Continuous Feedback Cycle** ▸ The C2PA Steering Committee should review feedback from other members, researchers, civil society organizations, and the public to continue improving the standard.

  To ensure interoperability and maintain trust, stakeholders should actively engage with the C2PA Conformance Program and leverage its Conformance Explorer to verify the status of generators, validators, and CAs. This alignment is critical for scaling adoption and for ensuring that provenance signals remain credible as the ecosystem evolves.
- **Red-Teaming and Analysis to Identify and Mitigate Weaknesses** ▸ MIA stakeholders should engage in ongoing technical and sociotechnical red-teaming and analysis to probe for weaknesses in the methods, to support transparent disclosure of strengths and weaknesses, and to guide refinements of technical approaches, policies, and laws. To support this, C2PA should promote ongoing intensive red-teaming and analysis of its specifications and implementations.
- **Iterative Policy Development** ▸ Policy efforts should drive adoption of technical methods for which there is implementation readiness, while building an understanding of limitations that may exist to inform the public's interpretation of provenance reliability. Policy expectations should be incrementally lifted in tandem with advancements in research and technical methods that can be deployed at scale.
- **Policy Accommodations** ▸ The report findings underscore the value of robust media integrity and authentication practices yet also reflect the reality that technical and operational contexts can vary widely. As the ecosystem evolves, it may be prudent for policy approaches to accommodate a range of implementation scenarios, ensuring that efforts to strengthen media authenticity remain effective and relevant across diverse environments.

# I. Introduction: Media Integrity and Authentication Methods

Recognizing the evolving complexities of media integrity and authentication methods, this report begins with an overview to establish a shared understanding of key terms used throughout this report. We build upon pre-existing taxonomies while providing additional context relevant to our specific findings.[3] A comprehensive glossary of all relevant terminology is also available in Appendix 1.

Three core technologies exist for authenticating audio-visual media: *provenance metadata*, *watermarking*, and *fingerprinting*. These technologies can be applied to authentic media (e.g., camera-captured), fully synthetic media (i.e., AI- generated), or mixed media (i.e., AI-modified). Each method serves a different purpose with varying levels of effectiveness. While this section defines several available MIA options, the report defaults to provenance (secure metadata), watermarks (imperceptible), and fingerprints (soft hashes).

While these technologies can also be applied to text, this analysis prioritizes images, audio, and video—modalities where these methods have been adopted at scale and that offer higher reliability for validation results.[4]

---

[3] See Bilva Chandra, Jesse Dunietz, Kathleen Roberts, Yooyoung Lee, Peter Fontana, and George Awad. Reducing risks posed by synthetic content an overview of technical approaches to digital content transparency, National Institute of Standards and Technology, 2024. https://doi.org/10.6028/NIST.AI.100-4 and Partnership on AI. Building a Glossary for Synthetic Media Transparency Methods, 2023. https://partnershiponai.org/resource/glossary-for-synthetic-media-transparency-methods-part-1/.

[4] Disclosing and detecting AI-generated or AI-modified text comes with a number of complexities and challenges. For one, secure metadata (per the C2PA standard) can only be applied to limited text scenarios – for generation within structures that support metadata such as Office documents and PDF files. While watermarking can be applied to text, robustness may not be high enough for many scenarios, resulting in high error rates during detection. For instance, all text watermarking methods are vulnerable to attacks (e.g., manual or tool-assisted paraphrasing attacks) and suffer performance degradations when used to detect short-form text. (See, e.g., Hanlin Zhang, Benjamin L. Edelman, Danilo Francati, Daniele Venturi, Giuseppe Ateniese, and Boaz Barak. Watermarks in the sand: Impossibility of strong watermarking for generative models. *arXiv preprint Xiv:2311.04378*, 2023 and Xia Han, Qi Li, Jianbing Ni, and Mohammad Zulkernine. Robustness Assessment and Enhancement of Text Watermarking for Google's SynthID. *arXiv preprint arXiv:2508.20228*, 2025.)

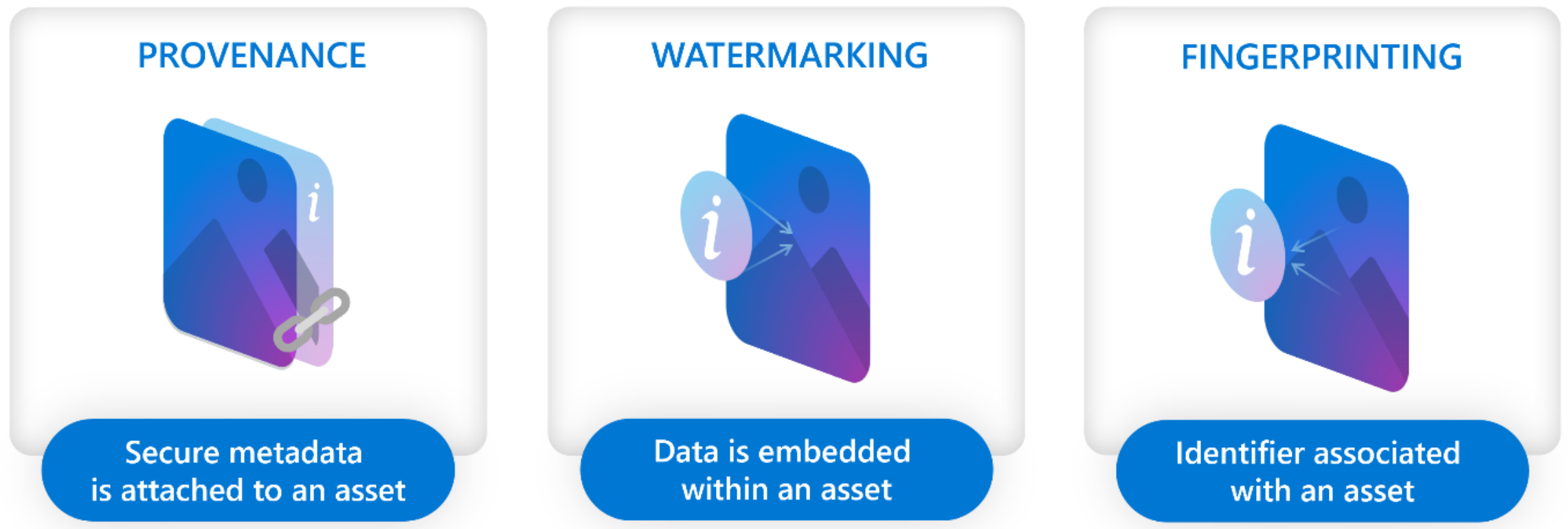

*Figure 1: Simplification of provenance, watermarking, and fingerprinting media integrity methods.*

### *Provenance Metadata*

The intention of provenance technology is to convey the source and history of digital content – details often referred to as metadata. The greatest area of innovation with provenance has been around secure provenance, which requires secure metadata.

With **secure provenance,** per the C2PA open standard, metadata is attached to the content file to communicate information about the content's origin and history, such as how it was made, whether it's been edited, and if so, where/how it was edited. This metadata is then *cryptographically signed* with a digital signature which offers a layer of protection. If the metadata is tampered with, the digital signature will be broken. The metadata that was included by the signer is made accessible via a manifest, which allows consumers to validate that the media asset is unchanged, and the signer is intact.

Using C2PA to signal the asset's signer unlocks additional benefits; authors and creators can be protected from impersonation and forgery, and recipients and relying parties can use the information to distinguish between trusted sources and untrusted sources like bots employing AI tools and other attackers. Importantly, validated provenance data is not proof that the content is true; trust in the content depends on the degree to which the consumer trusts the signer.

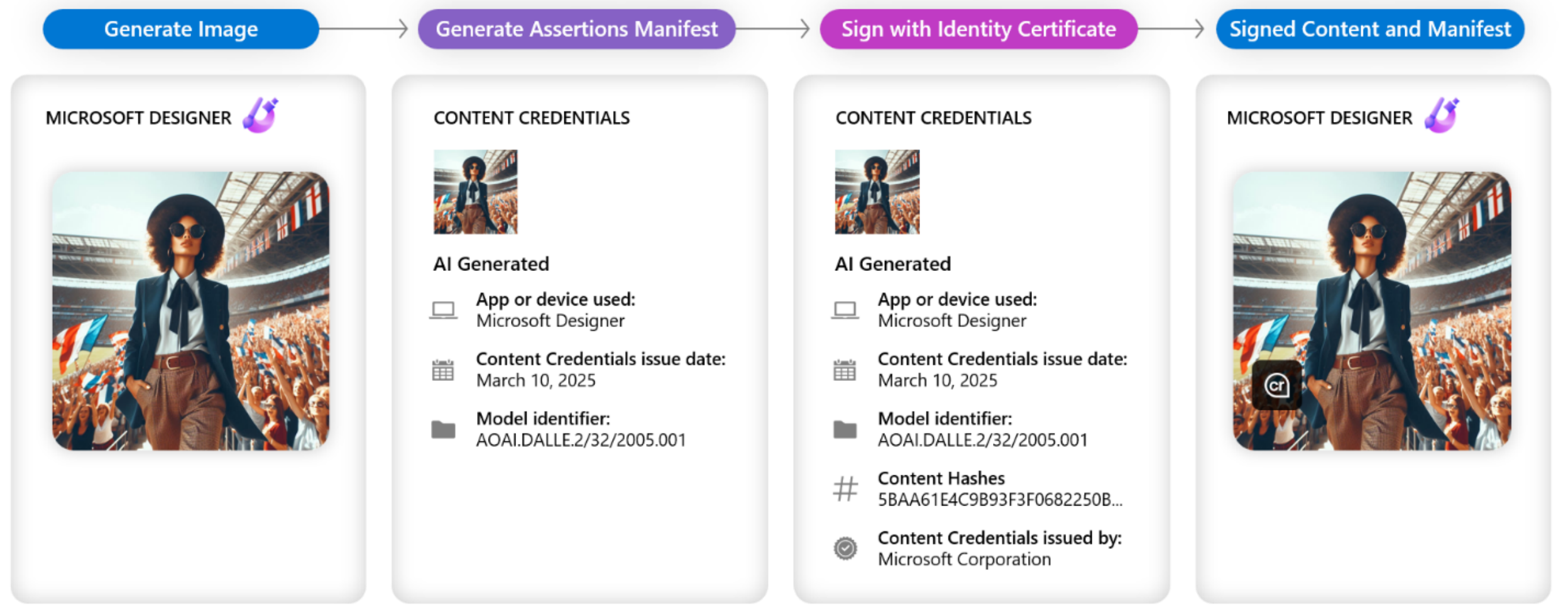

*Figure 2: End-to-end flow for adding provenance based on the C2PA standard. Cryptographically signed metadata can be automatically bound to AI generated or modified media (pictured above), media captured by a camera or recording device, or retroactively added to media with assertions the signer makes about the media's origin and history.*

The above innovations with secure provenance build on a long history of metadata use but without such security guarantees. In the case of **non-secure metadata**, metadata is similarly attached to the content file to communicate information about the content's provenance, but it is not cryptographically signed to protect the integrity of the information. Thus, non-secure metadata can be easily edited/manipulated. Many commonly adopted industry standards exist for such metadata—for example, Exif (Exchangeable Image File Format) metadata for photos or IPTC (International Press Telecommunications Council) metadata for images.

### *Watermarks*

A watermark is information embedded into a digital asset (e.g., image, audio, video) and can assist in verifying the authenticity of the content or characteristics of its provenance, modifications, or conveyance. There are two primary types of watermarks associated with MIA methods: imperceptible and perceptible.

**Imperceptible watermarks** are invisible or inaudible data embedded into the content of a media asset. Imperceptible watermarks involve subtle perturbations/modifications of the content that are hard for humans to detect. An encoder inserts an imperceptible watermark on a piece of content by slightly modifying its bytes, and a decoder extracts the watermark from the content even if the asset has been altered. Watermarks usually carry metadata information about provenance, or at a minimum, a reference/tracking ID that can be used to retrieve such information.

Imperceptible watermarks can be implemented in multiple ways, depending on the intended protection requirements. A *"fragile"* imperceptible watermark is designed to become invalid with mild changes to the content, while the *"robust"* imperceptible watermark method is designed to withstand certain types of attacks or modifications.

Availability of the watermarking algorithm and associated cryptographic information—essential for applying or validating a watermark—varies based on access design.

- A *"private"* watermark is accessible only to a limited group, such as internal employees.
- A *"restricted"* or *"controlled"* watermark involves licensed access to the algorithm or decoder, shared with a select set of trusted parties (e.g., journalists, social media platforms).
- A "public" watermark is either open source or paired with a decoder that's publicly available.

**A perceptible watermark** is visible or audible and easy to detect (e.g., logos, text or shapes overlayed on an image). Perceptible watermarks can be valuable in non-malicious use cases (e.g., for setting norms of disclosure). However, they are vulnerable to removal or forgery, reducing their protective value in adversarial cases and potentially contributing to confusion (for instance, if they are forged and added to media on which they do not belong).

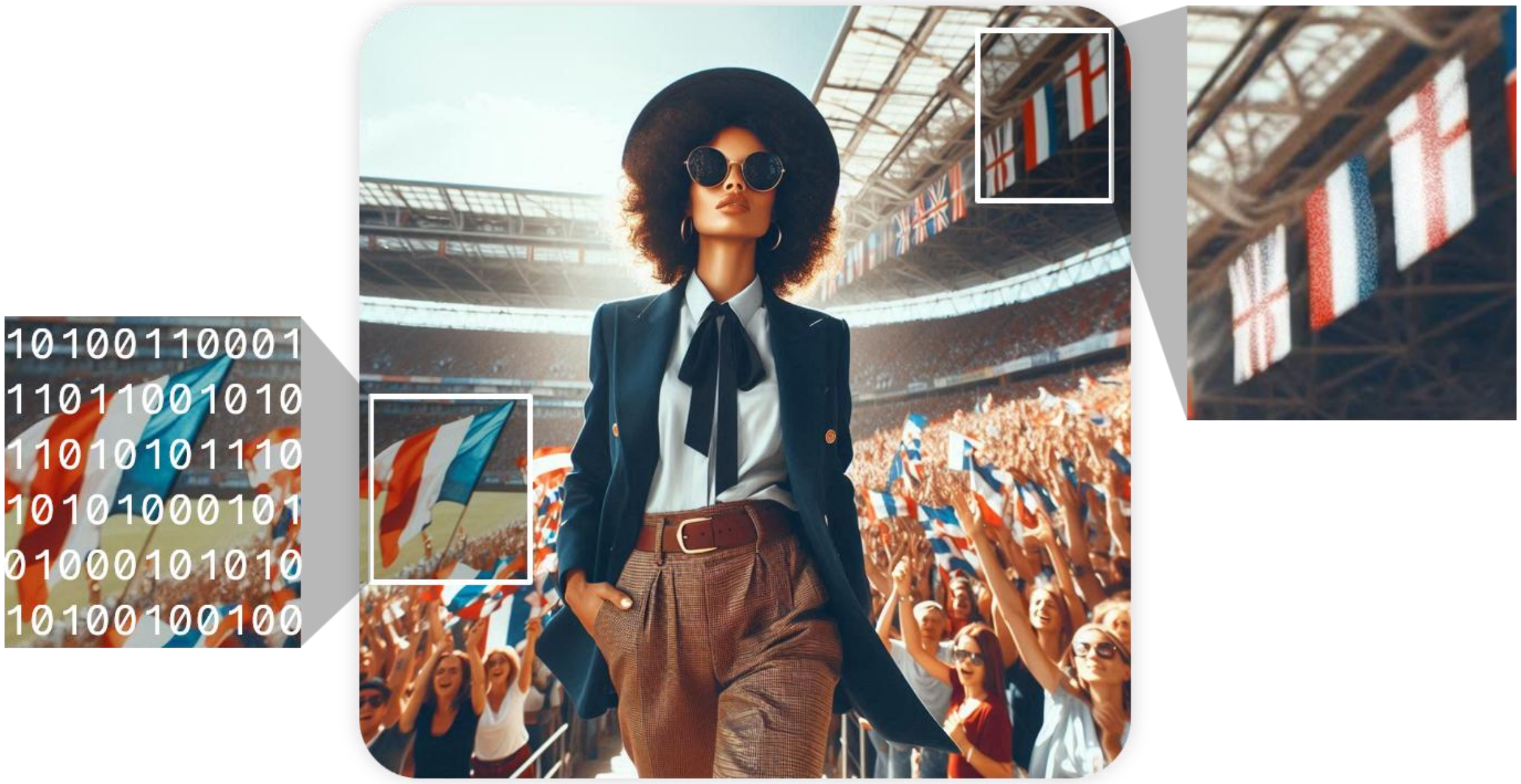

*Figure 3: Simulated rendering of imperceptible watermarks—made visible for illustrative purposes.*

### *Fingerprints*

A **fingerprint** is an identifier ("hash") computed from the media asset using an algorithm known as a hash function. This hash can then be compared to hashes stored in a database to see if a match is identified.

Fingerprints are generally most useful if the identifier is not changed when the asset is modified. The method is used to track unauthorized distribution or modification of media assets. In the context of online safety, hash matching is used to detect known harmful, illegal, and/or sensitive images and videos.

Fingerprinting methods include hard hashing (used to identify exact matches) or soft hashing (used to identify similar matches). This paper focuses on soft hashing, or perceptual hashing, that derives a small *soft* hash of the media content from a lower resolution / dimensionality projection of the content. The 'soft' nature of the hash ensures that minor editorial modifications still yield the same fingerprint. Perceptual hash functions can be grouped into three categories based on their underlying design principles: (1) dividing images into squares, (2) transforming images into waves, and (3) using machine learning models. Due to limitations such as hash collisions and attacks outlined in Appendix 2, manual review of potential matches is recommended, which can significantly increase storage costs.[5]

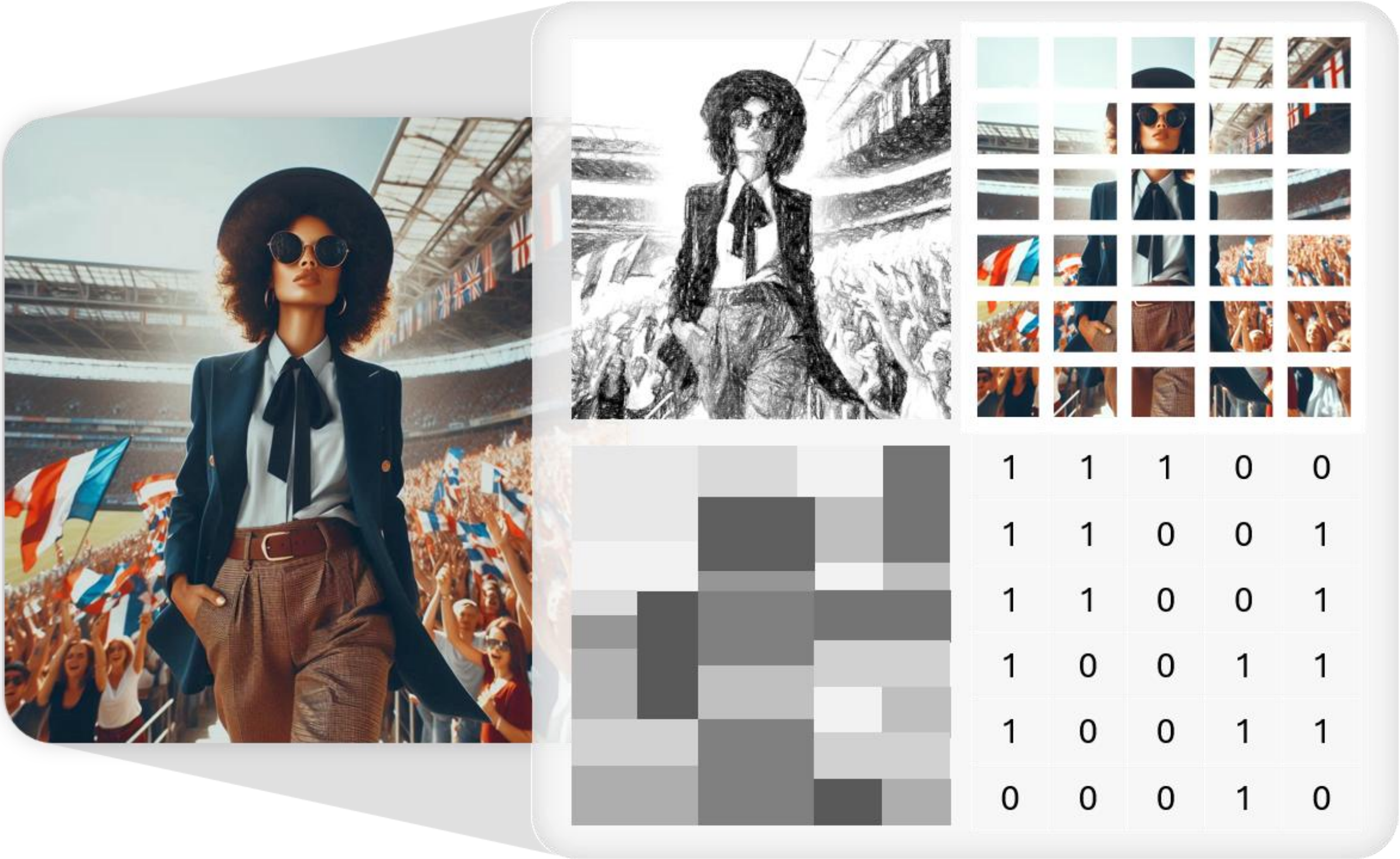

*Figure 4: Illustrative examples of soft hashing methods.*

[5] Storage costs vary across media type. For instance, storing a hash for a single image will be much lower than storing frame-by-frame hashes of a video file's imagery and audio track. Further, storing only the hash that corresponds to an image will be much lower than also storing a thumbnail representation of the image. As such, there are important trade-off considerations between storage costs and the ability to support manual forensics.

## II. Rising Need for Media Integrity and Authentication Methods

### *Evolving landscape of deceptive and harmful online content*

In 2022, *interactive* and *compositional* deepfakes were futuristic capabilities on the horizon.[6] But manipulations of authentic media and photorealistic AI generations[7] are becoming increasingly easy to produce and harder to distinguish. Advancements with AI capabilities are also paving the way for real-time engagement with hyper-realistic audio and video representations of any individual.[8] This presents new challenges for tackling fraud[9] and the rising scale of deceptive, harassing, and illegal content.[10] At the same time, the ecosystem is diverging on content policy approaches and shifting with less consensus on appropriate guardrails to mitigate content harms.

We anticipate further complexity ahead based on emerging and anticipated trends:

- Content will continue to move from being "purely authentic" or "purely synthetic" to a mixture of the two that evolves over the content lifecycle. As the bulk of media becomes a combination of real and synthetic, synthetic will eventually eclipse authentic media.
- We can expect to see GenAI systems interleaving authentic clips with synthetic clips and modifying video scenes with new capabilities and techniques for enhanced photorealism (e.g., modifying scenes through multi-camera views and vantage points).
- More and more content will be created offline on local devices, where media authentication methods can be difficult to secure and easy to hack.
- Compositional deepfakes will surface, as actors integrate observed, expected, and engineered world events over time to create persuasive, synthetic histories.
- Provenance use for authentic media will see an uptick, building on recent momentum from new camera and camcorder releases with built-in C2PA-based provenance.[11] The pendulum will shift back from an intense focus on synthetic content generated by AI to a focus on authentic content and validating what is real.

---

[6] Eric Horvitz. On the horizon: Interactive and compositional deepfakes. In *Proceedings of the 2022 international conference on multimodal interaction,* pp. 653-661, 2022. https://arxiv.org/abs/2209.01714
[7] Matt Growcoot. People Are Using Camera Filenames to Make Midjourney More Photorealistic, April 2025. https://petapixel.com/2025/04/07/people-are-using-camera-filenames-to-make-midjourney-more-photorealistic/
[8] See, e.g., Sicheng Xu, Guojun Chen, Yu-Xiao Guo, Jiaolong Yang, Chong Li, Zhenyu Zang, Yizhong Zhang, Xin Tong, and Baining Guo. Vasa-1: Lifelike audio-driven talking faces generated in real time. *Advances in Neural Information Processing Systems*, *37*, 660-684. 2024. https://arxiv.org/abs/2404.10667; ElevenLabs. Free AI Voice Generator & Voice Agents Platform, https://elevenlabs.io; GitHub. DeepFaceLive: Real-time face swap for PC streaming or video calls, https://github.com/iperov/DeepFaceLive.
[9] Victor Tangermann. OpenAI's New Image Generator Is Incredible for Creating Fraudulent Documents, April 2025. https://futurism.com/the-byte/openai-new-image-generator-fake-receipts
[10] Microsoft**.** Protecting the Public from Abusive AI-Generated Content, 2024. https://aka.ms/ProtectThePublic
[11] Recent releases with C2PA spec implementations include Google Pixel 10, Nikon Z6 III, Leica M11-P, v, Canon EOS R1 and EOS R5 Mark 2, among others.

These ecosystem changes in both AI capabilities and content policy approaches suggest challenges ahead with the public's ability to discern authentic representations of real-world events from synthesized and/or modified content.

Overall, the study committee asserts the importance of identifying and pursuing opportunities to strengthen understanding, adherence to, and advancement of content authenticity and provisions of *reliable* provenance information for *both authentic and synthetic content*. A priority in the world of rising quantities of AI-generated content must be certifying reality itself.

### *Media Integrity and Authentication Methods in Legislation*

Requirements to employ media integrity methods are increasingly surfacing in legislation. Although specifics differ across states and countries, requirements generally proposed by policymakers imply:

- C2PA provenance to include detailed data about generated or modified material and to employ a disclosure method consistent with industry standards.
- Watermarking to increase difficulty in removing the C2PA provenance.
- Provision of a provenance or watermarking validation tool and/or display of provenance information on online platforms such as social media sites.

Some recently proposed bills also call for provenance for authentic media; bills have included provisions requiring that state agencies add provenance to all media they publish or that recording or 'capture' devices (such as photography cameras, mobile phones with built-in cameras or microphones, and voice recorders) provide users the option to add provenance data.

The legislative landscape remains dynamic, but recently passed and proposed legislation with provenance requirements includes:

- **Digital Services Act [passed – *in effect*]** requires companies to identify and mitigate systemic risks with respect to AI, including the generation of deceptive content. The EU Commission specifically identified provenance as a key mitigation in the context of its election guidance.
- **China Regulation on the Management of Deep Synthesis of Internet Information Services [passed – *in effect*]** requires covered "deep synthesis service providers" to attach symbols to AI generated or edited content, store log information, apply conspicuous labels to content that may confuse or mislead the public, and provide and notify users of the ability to provide prominent labels. It also prohibits organizations and individuals from deleting, altering, or concealing such labels.
- **California AI Transparency Act [passed, amendment process ongoing– *in effect in August 2026*]** currently includes requirements for providers of generative AI systems, large online platforms, capture devices, and platforms that host generative AI systems. Beginning August 2026, providers of generative AI systems will be required to offer users the option to include a manifest (i.e., perceptible) AI-generated content disclosure and to attach a latent (i.e., hidden) disclosure in all AI-generated or altered content, containing details like the provider's name, AI system version, and content creation date. Both types of disclosure must be "permanent or extraordinarily difficult to remove" to the extent technically feasible. Providers must also make a tool to detect content from their systems (i.e., a C2PA provenance and/or watermarking validation tool) publicly accessible.
Beginning January 2027, large online platforms will be required to (1) detect whether any provenance data that is compliant with widely adopted specifications adopted by an established standards-setting body is embedded into or attached to content distributed on the platform; (2) to the extent technically feasible, retain any system provenance data or digital signature that is compliant with such specifications; (3) provide a user interface to disclose system provenance data that reliably indicates the content was generated or substantially altered by a GenAI system or captured by a capture device; and (4) allow users to inspect all such available system provenance data in an easily accessible manner.

Also starting January 2027, generative AI system hosting platforms will be prohibited from knowingly making available a GenAI system that does not place disclosures that are permanent or extraordinarily difficult to remove into content created or substantially modified by the GenAI system. Starting January 1, 2028, manufacturers of capture devices that can record photographs, audio, or video content and are produced for sale in California on or after that date will be required to provide users the option to include certain provenance data in the user's captured content via a latent disclosure.

- **EU AI Act [passed – *in effect, August 2026*]** requires providers of generative AI systems to design their systems in such a way that synthetic audio, video, text and image content is marked in a machine-readable format, and detectable as AI-generated or manipulated. Requirements apply as far as technically feasible, considering specificities and limitations of different types of content and the generally acknowledged state-of-the-art, as may be reflected in relevant technical standards. A Code of Practice is being developed to define a path to compliance with requirements. Failure to comply can result in fines of up to 3% of global revenue or up to €15M, whichever is higher.
- **India Information Technology (Intermediary Guidelines and Digital Media Ethics Code) Amendment Rules, 2025 [proposed]** would require providers of computer resources used to generate synthetic information to prominently label such content or embed within it a permanent unique metadata or identifier that covers 10 percent of the content's surface area or duration, while prohibiting the providers from enabling the disclosure's modification or removal. In addition, significant social media platforms would be required to ask users to declare whether information they upload is synthetically generated, use technical measures to verify the accuracy of the declaration, and ensure that content confirmed to be synthetically generated (via a declaration or technical measures) is labeled as such.

- **Korea AI Basic Act [passed – *in effect, January 2026*]** requires generative AI products and services to indicate that results were generated by generative AI. It further requires that for generative AI outputs that are difficult to distinguish from reality — such as voices, images, or videos — providers clearly inform users that the results have been created by an AI system, with certain exceptions.

Legislation has passed calling for measures that are not technically feasible, with the aspiration that perceptible and imperceptible disclosures be "permanent or extraordinarily difficult to remove."[12] In other cases, provisions may inadvertently dilute the quality of provenance information displayed to content consumers. For instance, unintended consequences may stem from requirements for capture devices to include disclosures, and for platforms to detect and make available *any* provenance data, that are "compliant with widely adopted specifications adopted by an established standards-setting body."[13] Such requirements may result in broad consumption of insecure provenance information (such as IPTC or EXIF metadata) that has been manipulated. Further, requirements to add perceptible watermarks may cause confusion in cases of forgery or discourage people from consulting high-confidence provenance information via a validation tool, if such perceptible disclosures are taken at face value. As new legislation is proposed, and code of practice guidance is shaped for current legislation, it will be important for policymakers to understand:

- The state of the art of all media integrity technologies, as it relates to their security, robustness, reliability, and limitations.
- The importance of differentiating between secure and insecure provenance information.
- The importance of prioritizing common and interoperable approaches for consistency across the ecosystem, while also maintaining flexibility to adapt to the evolving state of the art and employ the most appropriate media integrity method for edge cases. Potential edge cases include scenarios not yet supported by secure provenance (e.g., new modalities for which the C2PA standard has not yet been extended); cases where there is a lack of robust techniques for watermarking (e.g., black and white images); or cases where low-security environments may make open-source technologies more appropriate to deploy than protected technologies (e.g., proprietary fingerprinting) in order to mitigate reverse engineering or misuse of the MIA technologies themselves.

[12] The California AI Transparency Act currently requires that both perceptible and imperceptible disclosure be "permanent or extraordinarily difficult to remove" to the extent technically feasible. Perceptible disclosures can be easily removed by novice actors (see, e.g., Umar Shakir. Google's Gemini AI is really good at watermark removal, March 2025. https://www.theverge.com/news/631203/google-gemini-flash-2-native-image-generation-watermark-removal and Slashdot. Sora 2 Watermark Removers Flood the Web, October 2025. https://tech.slashdot.org/story/25/10/07/2110246/sora-2-watermark-removers-flood-the-web), while imperceptible disclosures with state-of-the-art robustness will still be removable by sophisticated actors. Recent research demonstrates methods that achieve near-perfect watermark removal with minimal degradation to image and audio quality (see e.g., Fahad Shamshad, Tameem Bakr, Yahia Salaheldin Shaaban, Noor Hazim Hussein, Karthik Nandakumar, and Nils Lukas. First-Place Solution to NeurIPS 2024 Invisible Watermark Removal Challenge. The 1st Workshop on GenAI Watermarking, collocated with ICLR 2025. https://openreview.net/forum?id=wLaP37BrhE and Patrick O'Reilly, Zeyu Jin, Jiaqi Su, and Bryan Pardo. Deep Audio Watermarks are Shallow: Limitations of Post-Hoc Watermarking Techniques for Speech. The 1st Workshop on GenAI Watermarking, collocated with ICLR 2025. https://openreview.net/forum?id=44TCZ5XTuR), and that diffusion-based image editing can effectively break state-of-the-art robust watermarks designed to withstand conventional distortions (see Wenkai Fu, Finn Carter, Yue Wang, Emily Davis, and Bo Zhang. Diffusion-Based Image Editing: An Unforeseen Adversary to Robust Invisible Watermarks. *arXiv preprint arXiv:2511.05598*, 2025.).
[13] The California AI Transparency Act currently includes such requirements.

- The importance of considering privacy as a critical component of provenance legislation, while acknowledging the value that identity can play in provenance, *if* an individual or organization deliberately includes such information.
- The benefits of granular provenance information, where practical, given challenges with otherwise determining if edits made were material.

## III. Identifying Limitations and Attack Vulnerabilities

### *Limitations of today's discrete methods*

Despite rising calls for disclosure methods that are *permanent* and *robust to attacks*, and despite the respective benefits of each currently available approach, no foolproof method for media integrity and authentication exists. Using these methods individually presents a host of issues; each method may fail to return results, or a method may return misleading results if it is solely relied upon for authentication.

### *Technical Attacks and Sociotechnical Attacks Drive Confusion*

Cryptographically signed metadata (i.e., C2PA manifests), watermarking, and fingerprinting are all vulnerable to attacks. Such attacks (summarized in *Figure 5* and outlined in detail in Appendix 2) can result in erroneous information being displayed about the media's provenance during the authentication process. These faulty or misleading results may be served to the public, an organization's employees, partners and/or customers. Attacks fall into key categories including:

1. Misattribution or mischaracterization: adding/modifying a C2PA manifest or watermark or modifying the media content to change its fingerprint to (a) make it look like an asset was created by an entity when it wasn't or (b) make the asset appear synthetic when it is truly authentic or vice versa. (This might be done, for instance, for reputational harm, illegal purposes, or to spread disinformation.)
2. Removal: removing a C2PA manifest or watermark or modifying the media content to change its fingerprint to be able to use the asset without restrictions or to spread uncertainty about its provenance.
3. Denial of service: adding/modifying a C2PA manifest or watermark and/or modifying the media content to change its fingerprint to overload or make a validation service unavailable. (Such an attack could be used as a precursor to a time-sensitive disinformation campaign.)

DEFENSE IN DEPTH

| State-of-the-Art* (SOTA) | BENEFITS | | | | for TECHNICAL ATTACKS | | | | for SOCIOTECHNICAL ATTACKS | | |
|---|---|---|---|---|---|---|---|---|---|---|---|
| | Industry Standard | Digitally Signed | Detection Type | Forensics Support | Forgery Resistant | Resistant to Platform Removal | Resistant to Removal Attacks | Always Available | High Confidence Results *w/ Provenance Display* | Low Confidence Results *w/o Provenance Display* | Misleading Results Mitigation |
| PROVENANCE | ✓ | ✓ | Exact | ✓ | ✓ | | | | ✓ | | ✓ |
| WATERMARK | | | Sensitive to Noise | ✓ | | ✓ | | | | ✓ | |
| FINGERPRINT | | | Sensitive to Noise | ✓ | ✓ | N/A | N/A | ✓ | | ✓ | |
| PROVENANCE + WATERMARK | ✓ | ✓ | Less Sensitive to Noise | ✓+ | ✓ | ✓ | | | ✓ | ✓ | ✓ |
| PROVENANCE + WATERMARK + FINGERPRINT | ✓ | ✓ | Much Less Sensitive to Noise | ✓++ | ✓ | ✓ | | ✓ | ✓ | ✓ | ✓ |
| DISPLAY (UX) | | | | | | | | | | | ✓ |

**State-of-the-Art methods refer to provenance (secure metadata), watermarks (imperceptible), and fingerprints (soft hash)*

*Figure 5 This table presents a comparative snapshot of media integrity methods, highlighting capabilities and limitations of individual approaches, method combinations, and exposure to technical and sociotechnical attacks.*

## IV. Pursuing High Confidence Results

The benefits of leveraging multiple MIA technologies have been widely discussed given their respective capabilities and limitations (see Appendix 2). However, less research has focused on how to optimally combine these technologies and how doing so may improve validation results.[14] As part of this study, we explore combinations of and links between these technologies to determine optimal validation results in light of an extensive set of modifications and attacks media may undergo.

### Goal

C2PA manifests can be reinforced with watermarking and fingerprinting, enabling media to be authenticated even when the C2PA manifest has been stripped from the file.[15] This can support provenance lookup for the public, for incident response purposes (in cases of unsophisticated adversaries/hobbyists[16]), and to support compliance with legislative requirements for difficult to remove provenance. Beyond manifest recovery, we also want to ensure media is not authenticated if the asset has been tampered with after provenance information was added or if the manifest or associated provenance information has been intentionally manipulated to convey inaccurate information.

---

[14] For one such exploration of this topic, see John Collomosse and Andy Parsons. To Authenticity, and Beyond! Building Safe and Fair Generative AI Upon the Three Pillars of Provenance. In *IEEE Computer Graphics and Applications*, vol. 44, no. 3, pp. 82-90, May-June 2024, doi: 10.1109/MCG.2024.3380168.
[15] Ibid.
[16] We expect advanced adversarial attacks (e.g., those of nation state actors) will be able to undermine all three MIA techniques, even when the algorithms/implementations used are state of the art.

## Approach

We explore which **approach for authentication yields the highest confidence results possible while avoiding the risk of displaying provenance information in ways that could mislead or confuse** the public. In doing so, this report intends to lay the technical groundwork for effective content provenance. We leave to follow-on work in-depth studies and monitoring of issues and opportunities related to end-user experiences, alternative UX designs, and deeper, sociotechnical influences of rising uses, non-uses, and abuses of provenance technologies.

## Results: *High Confidence Authentication*

Considering known attacks on C2PA manifests, watermarks, and fingerprints, we create an extensive list of scenarios that are possible if the media is generated and signed with provenance information online (i.e., in a high-security cloud environment), all three media integrity methods are employed, and cross-referencing exists between the three methods (i.e., there is a database storing the manifest for each media file, a database storing the watermark reference ID for the media file that indexes to the associated C2PA manifest, and a database storing the fingerprints [soft hashes] computed for each media file that index to the associated C2PA manifest as well). We identify 60 unique combinations (where combinations are the set of potential validation results across C2PA manifest, watermark, and fingerprint computation that are possible given potential attacks and errors[17]) and five potential validation result states: invalid, indeterminate, presentation[18] may validate, presentation matches, and media validates. (See Appendix 3 for a breakdown of these combinations.)

We find 20 combinations[19] that map to only two scenarios that result in **high confidence that the media file is the same as the one that was signed and that the provenance assertions are as provided by the signer and unchanged.[20]** Using Microsoft as the signer, if the media generation and signing both occurred online, we can further affirm that those assertions are accurate.

---

[17] For instance, there are 3 potential validation results for C2PA (C2PA manifest is present and hashes match; C2PA manifest is present but hashes do not match; no C2PA manifest present). There are 5 potential validation states for watermarks (watermark detectable – and C2PA hashes match; watermark detectable – but C2PA hashes do not match; watermark detectable – but C2PA manifest is missing from the registry; no access [validator error]; watermark not detectable). And there are 5 potential validation states for fingerprints (fingerprint valid – and C2PA hashes match; fingerprint valid – but C2PA hashes do not match; fingerprint valid – but C2PA manifest is missing; no access [validator error]; fingerprint invalid).

[18] Here we use presentation to refer to the rendering of the media matching the C2PA hash on file for the media in the manifest store.

[19] See Appendix 3.

[20] It is the cryptographic verification and hard hash matching enabled by C2PA that allows us to have full confidence. Importantly, watermarking alone, without the use of C2PA, cannot provide high confidence that the media file is an exact copy of the one to which provenance information was added. With the C2PA hash verification, we can then check that the provenance information conveyed is unchanged.

1) The C2PA manifest is present and validated[21], including validation that the C2PA hashes match (i.e., the hash found in the manifest attached to the media and the hash stored on the server match) or

2) The watermark is detected, valid, and points to a C2PA manifest (in the manifest store) that includes a C2PA hash that matches the C2PA hash computed for the media.

## Results: *Low Confidence Authentication*

Our approach also surfaced more cases where provenance validation yielded low confidence or could not be asserted. For instance, there will be cases where a watermark is recovered, but the hashes do not match—suggesting the content was edited. In such cases, we cannot validate with high confidence and should avoid sharing results with users who might place undue trust in them. While low-confidence outcomes aren't suitable for consumer-facing use, lower-confidence signals (e.g., potential fingerprint matches) are valuable for certain types of scenarios, offering another layer of internal protection that forensics teams can leverage.

The flowgraphs below illustrate the potential for high-confidence (Figure 6) and low confidence (Figure 7) validation results and the sequential decisions that are made in authenticating the media. For example, if a C2PA manifest is present and valid and the hash matches, then the file is valid and there is no need to check watermark or fingerprint. For a holistic view of all five potential validation results, see Appendix 4.

---

[21] C2PA validation includes multiple steps: verifying that the media contains a stored, signed manifest; validating the signature (i.e., verifying the manifest has been signed by a party with a valid signing certificate that appears on a trust list); and comparing the hard hash of the asset itself with the hash included in the manifest (to verify that the media hasn't been tampered with since signing).

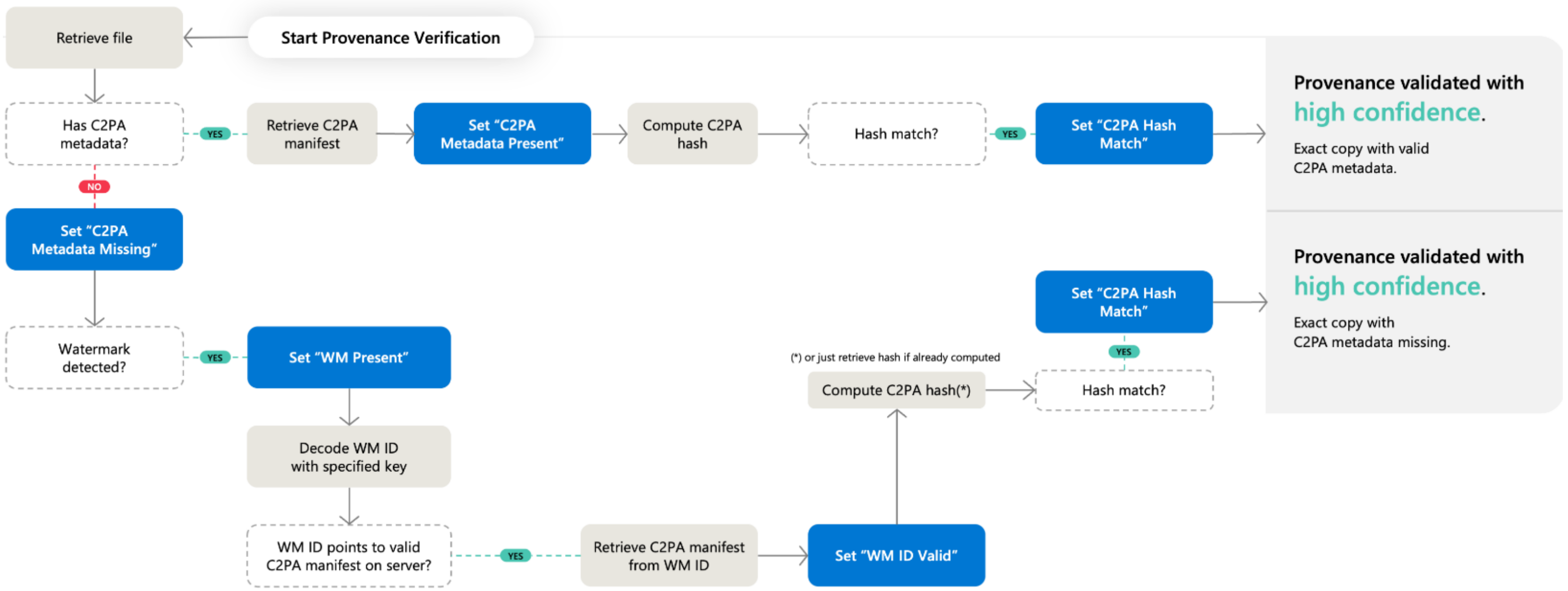

*Figure 5: The pathway toward high confidence results where green indicates the media is validated and media content matches the original copy to which provenance information was added by Microsoft, with the bar for high confidence being a cryptographically secure, validated C2PA manifest.*

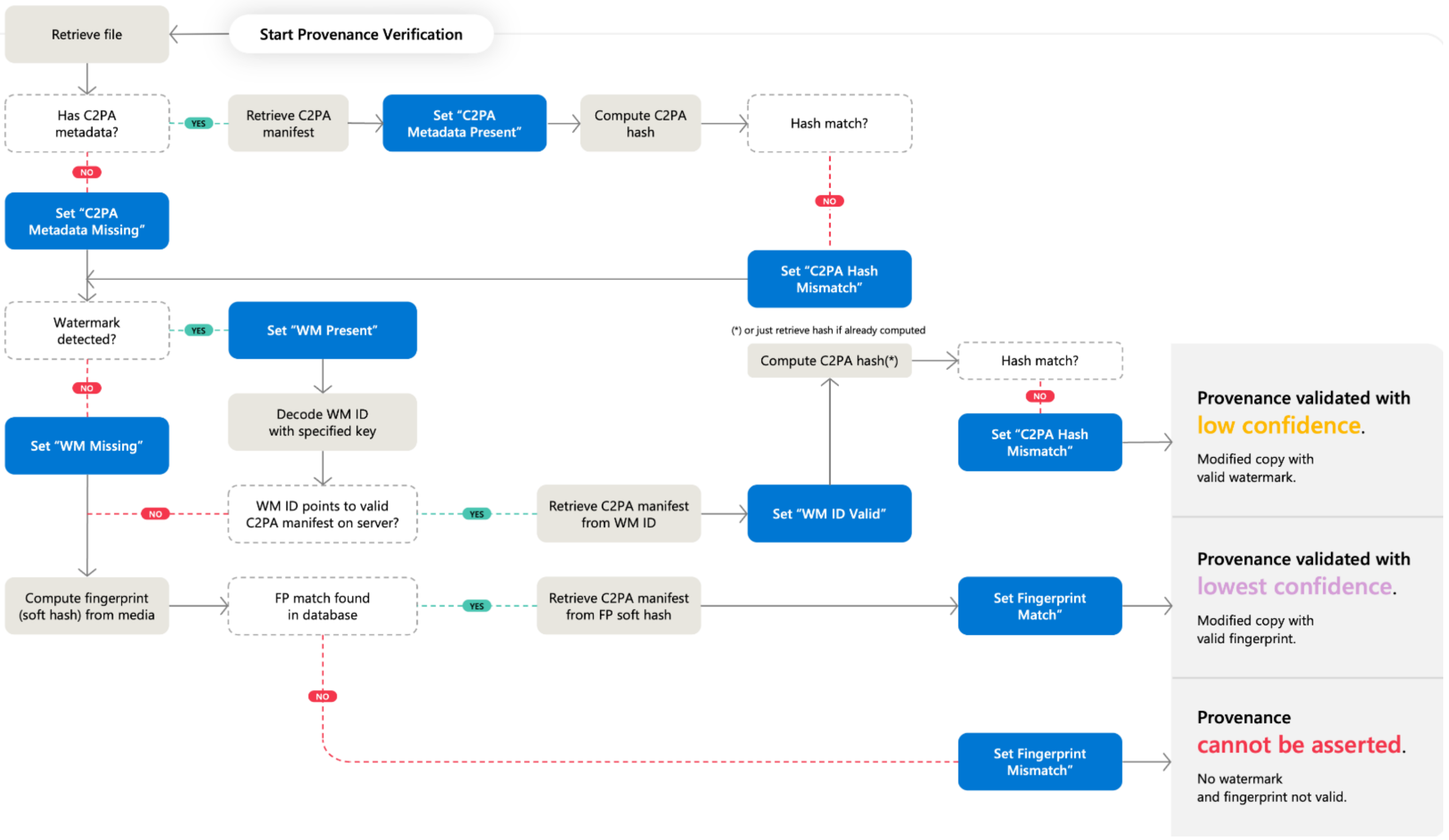

*Figure 6: The pathway toward low, lower, and no confidence results where yellow indicates media provenance cannot be asserted with high confidence, as the media content has been modified. Purple indicates the lowest level of confidence, as metadata is not present, the watermark cannot be recovered, but the fingerprint does match a provenance database entry. Red indicates certified media has been attacked beyond certification, or media that was never certified during production. We note that not all paths are equally likely; the red path will be the path for most files until C2PA is more prevalent.*

## Directions for High-Confidence Authentication

1. **Synthetic and Mixed Media** ▶ GenAI system providers should consider prioritizing provenance and watermarking for provenance recovery, where possible[22], for synthetic media generation and editing scenarios to enable high-confidence validation. To address cases involving heightened risk of abuse, organizations can explore provenance, watermarking, and fingerprinting to enable sequential authentication as needed.

2. **Authentic Media** ▶ Organizations should recognize and explore uses of provenance for certifying and raising trust in authentic content and records (such as photos, transcripts, documents), including uses of provenance to capture history of changes made through editing and post-production.

3. **Validation Tools** ▶ To minimize confusion and overreliance, we recommend provenance validation tool providers consider displaying only high-confidence results to the public. C2PA manifest validation and display should be the default way by which provenance information is shown on distribution platforms (e.g., social media sites) and publicly available first-party validation tools. Lower-confidence provenance results, if displayed, must be clearly distinguished from high-confidence indicators.

   *Additional considerations for validation tools:*

   - In some cases, validation tools may serve an important role in relaying *any* signals about media authenticity whether low or high confidence[23] to select audiences. Validation tool providers will need to weigh audience needs and media literacy, evaluate use cases, and make trade-off decisions about displaying less content that's highly reliable or more content that's less reliable, including the potential for adversarial provenance-style signaling or reporting.

   - Over time, as more parties adopt and display C2PA (e.g., media editing tools, newsrooms, social media and messaging platforms, web browsers), we expect to be able to validate more content and that most media validation will be able to be validated with high confidence. As provenance becomes more prevalent and content consumers begin to encounter media with provenance more frequently, they may grow increasingly skeptical of the authenticity of content lacking provenance. Should such a shift occur, validation tools may then want to consider also showing

---

[22] Maintaining flexibility will be necessary based on the scenario at hand. While prioritizing provenance supports high-confidence validation, there may be cases where provenance specifications (e.g., per the C2PA standard) have not been extended to account for use on new modalities. In other scenarios, watermarking may not be an effective solution. For instance, watermarking binary (black and white) images is also an evolving area, with a lack of robust techniques.

[23] See for example, WITNESS. Deepfakes Rapid Response Force – Technology Threats Opportunities, https://www.gen-ai.witness.org/deepfakes-rapid-response-force/.

Audiences such as journalists and civil society organizations may use validation tool outputs as only one part of a broader in-depth analysis, thus reducing overreliance on low-confidence outputs.

*provenance cannot be asserted* results. Other contextual or forensic evidence may be required in these situations to make inferences about the veracity of the content. The latter may be important, for example, in citizen capture of human rights abuses or atrocities, when formal provenance tools are not available and in other legitimate situations. [24]

4. **Accounting for Exceptions** ▸ As the use of secure provenance, for high-confidence results, won't be possible in all cases, industry should promote continued research and alignment on *display choices and media literacy*, to help mitigate legitimate, authentic media without provenance being discredited.

5. **Forensic Access** ▸ Companies should consider making MIA services available for forensic investigators to access lower-confidence provenance signals that are not suitable for general public display.

6. **Additional Safeguards** ▸ Due to security risks like potential "oracle attacks" on decoders, additional safeguards, such as employing multiple watermarks or unique keys, are necessary before making watermark detector tools publicly accessible.

## V. Stress Testing Authentication Results with Illustrative, Sociotechnical Attacks

Trust in the provenance assertions depends on adherence to the specification and the level of assurance met by the implementation; higher levels can be reached, for example, using secured and isolated cloud signing. Even if C2PA validation confirms that the content has not been manipulated post-signing, the manifest may still contain arbitrary attestations—either originally signed with the media or added later through re-signing. In addition to attacks where arbitrary provenance information is added, *there are also cases where provenance information is technically accurate but may be misleading depending on how it is displayed*. To illustrate this, we include exemplary attack scenarios that examine how misleading provenance information might be displayed to the public and explore potential mitigations.

### *Misleading the Public and Driving Widespread Confusion*

The true impact of attacks on MIA methods is most felt during the user experience when authentic media is faked as synthetic, AI-generated media is deemed authentic, or consequential details about the media's history are mispresented.

---

[24] See for example, WITNESS. Tomorrow's Great Digital Divide: Content With or Without Provenance, March 2025. https://blog.witness.org/2025/03/tomorrows-great-digital-divide/

## ATTACK SCENERIO 1: Authentic Faked as AI

**High Confidence Validation/Display Experience.** *Figure 8* illustrates the high confidence experience for a feasible attack whereby the attacker inputs an **(A)** authentic camera-generated image into a GenAI tool. **(B)** The attacker uses a generative AI "fill" or "erase" feature to make a subtle, insignificant edit. **(C)** The image is signed with secure provenance, accurately indicating that an AI tool was used to modify part of the image. With high-confidence validation, **(D)** both the watermark and associated C2PA manifest would be read. **(E)** The validator displays helpful context such as the thumbnail of the original image and the region where edits occurred. **(F)** Additional context allows the user to assess the materiality of the edits, thereby mitigating the attack.

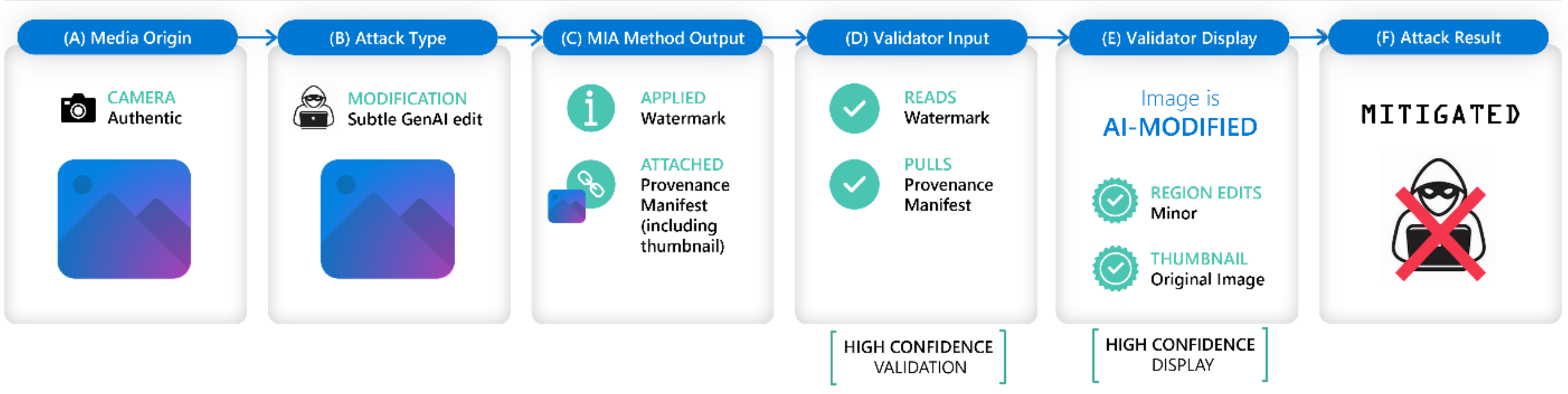

*Figure 7: Mitigating social attacks with high confidence results that provide additional context.*

**Low Confidence Validation/Display Experience**. In contrast (where steps **(A)**, **(B)**, and **(C)** are identical to *Figure 8*), a low-confidence validator like in *Figure 9* might simply **(D)** read the watermark and **(E)** display the authentic image is synthetic, AI generated, or modified, making it difficult for the user to **(F)** evaluate the media's origin and how, where, and the degree to which AI was used.

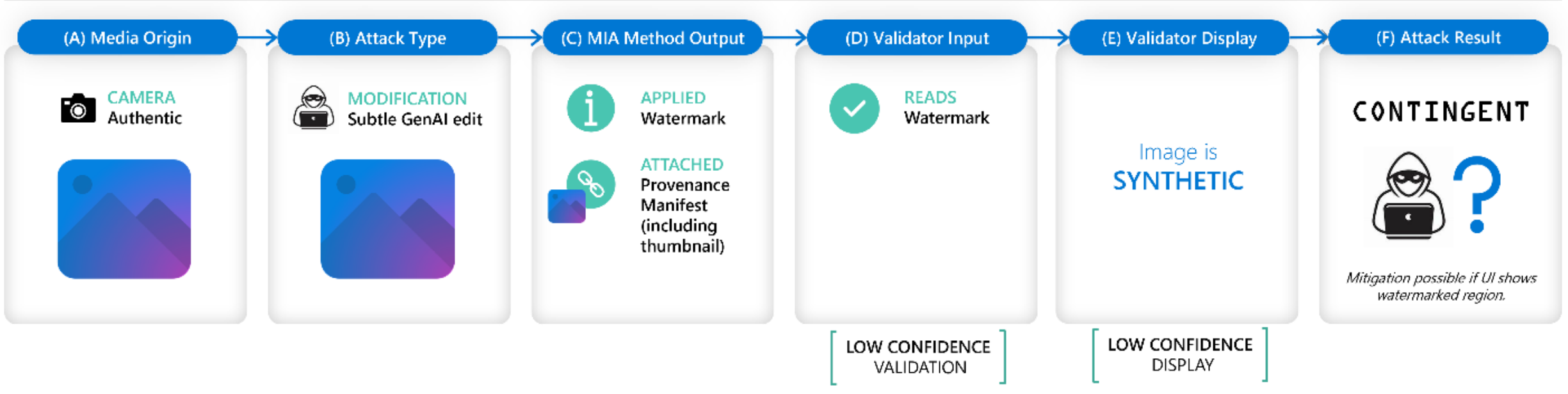

*Figure 8: Illustration demonstrating how low confidence validation and limited display context can confuse and mislead.*

Rising trends, including the use of precise and often subtle inpainting[25], as well as the use of AI verification/detection results to dismiss authentic content,[26] further point to the need for high-confidence validation and display experiences.

### ATTACK SCENERIO 2: AI Faked as Authentic

**High Confidence Authentication/Display Experience**. *Figure 10* captures a potential attack whereby an attacker **(A)** creates an AI-generated image and then **(B)** strips the C2PA manifest and watermark. With an intent to deceive, the attacker **(C)** adds a manifest with a camera-captured assertion to make the synthetic media appear authentic. This could be done by taking a screen capture of the image, signing it with another valid certificate that was stolen from a local device, and adding an assertion that it was camera-captured. A high confidence validator with a reliable list of trusted manifest signers will note **(D)** issues authenticating the media (e.g., if the certificate's theft was known and reported). The display **(E)** will indicate that results cannot be displayed or share a low security result to alert users and **(F)** mitigate the attack.

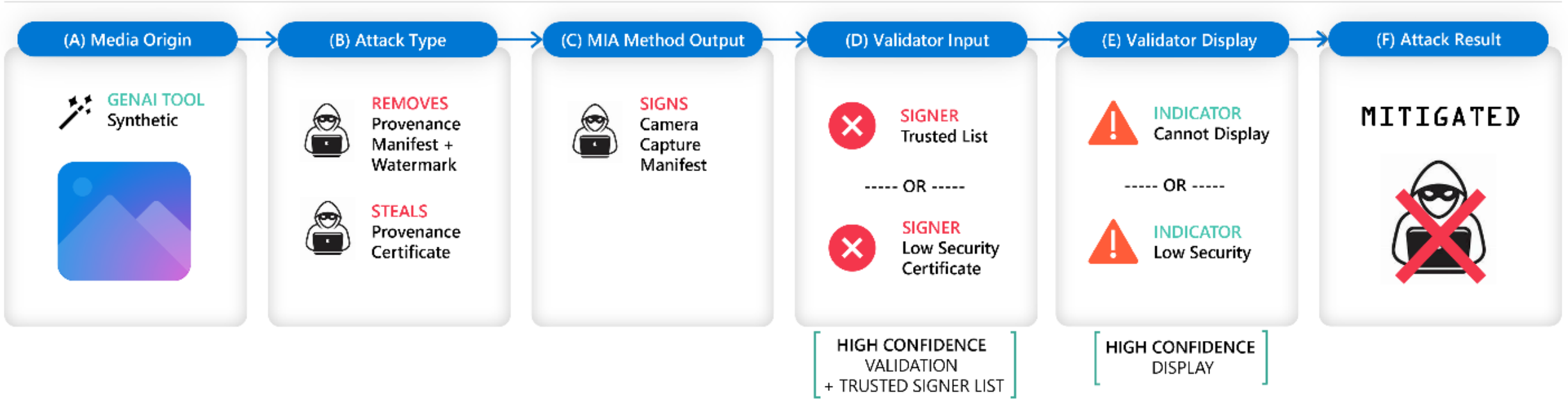

*Figure 10: The high confidence verification and display experience for AI displayed as authentic attack scenarios.*

**Low Confidence Authentication/Display Experience**. As *Figure 11* illustrates, even a high-confidence validator may display provenance information if the signer is listed as conforming to the C2PA specification and meeting reasonable security assurances. Refer to direction 4 as a potential path forward.

[25] See Zuzanna Wojciak and shirin anlen. Five Things 2025 Taught Us About AI Deception and Detection. TechPolicy Press, December 2025. https://www.techpolicy.press/five-things-2025-taught-us-about-ai-deception-and-detection/

[26] See Mahsa Alimardani. How Doubt Became a Weapon in Iran: AI manipulation, and the very suspicion of it, serves those who have the most to hide, January 2026. https://www.theatlantic.com/international/2026/01/iran-disinformation-ai-protests-doubt/685608/

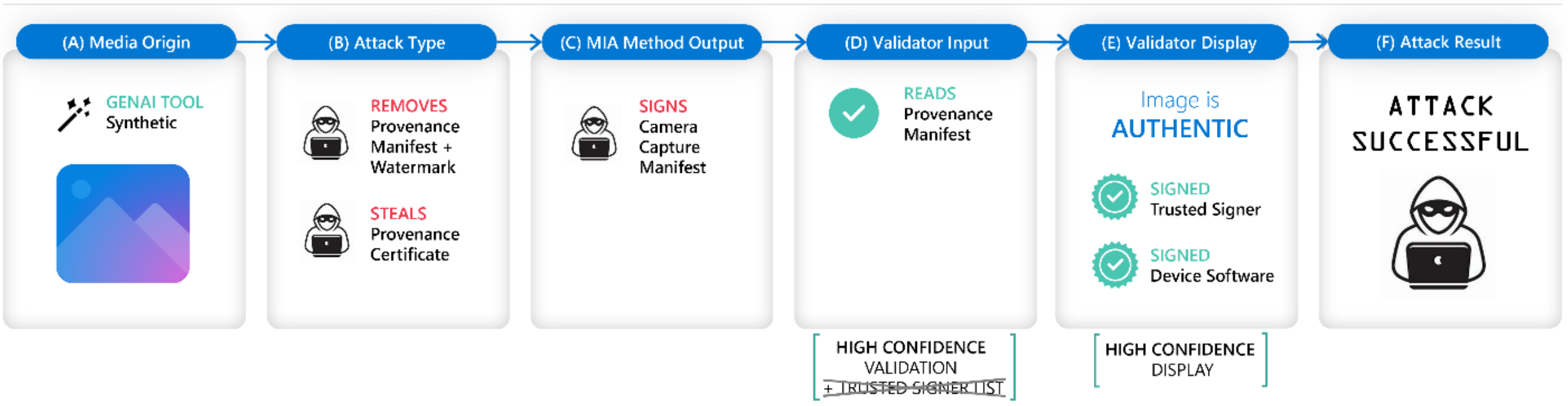

*Figure 11: Demonstrates how high confidence results can fail without a reliable list of trusted manifest signers and.or adeqaute consideration of/representation of signing certificate security levels.*

### ATTACK SCENARIO 3: Manipulated Metadata

**High Confidence Authentication/Display Experience**. Altering metadata including timestamp information can be especially consequential if an authentic image depicting a current event is misrepresented as a past occurrence with intent to deceive or when a past catastrophe is reframed as unfolding in the present to provoke alarm. In *Figure 12*, **(A)** an image was taken with (**B**) insecure provenance metadata added by the device upon camera-capture. An attacker then **(C)** manipulates the metadata to change the date and time of capture. Because a high-confidence validator only validates watermark and secure provenance metadata, which was **(D)** missing from this image, **(E)** the inaccurate metadata is not displayed, thus **(F)** mitigating the attack.

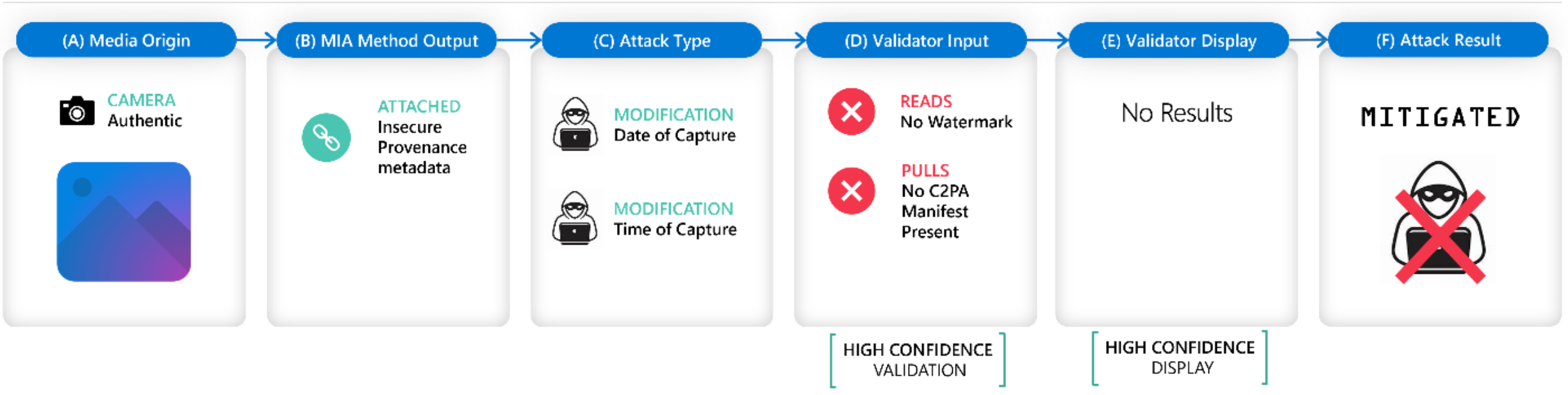

*Figure 12: The high confidence experience when an attacker manipulates the metadata.*

**Low Confidence Authentication/Display Experience**. *Figure 13* illustrates that this type of attack would be feasible if insecure provenance data was added by the initial camera (e.g., EXIF metadata) rather than secure provenance data (per the C2PA standard) and then read and displayed by the validator. See direction 5 as a potential path forward.

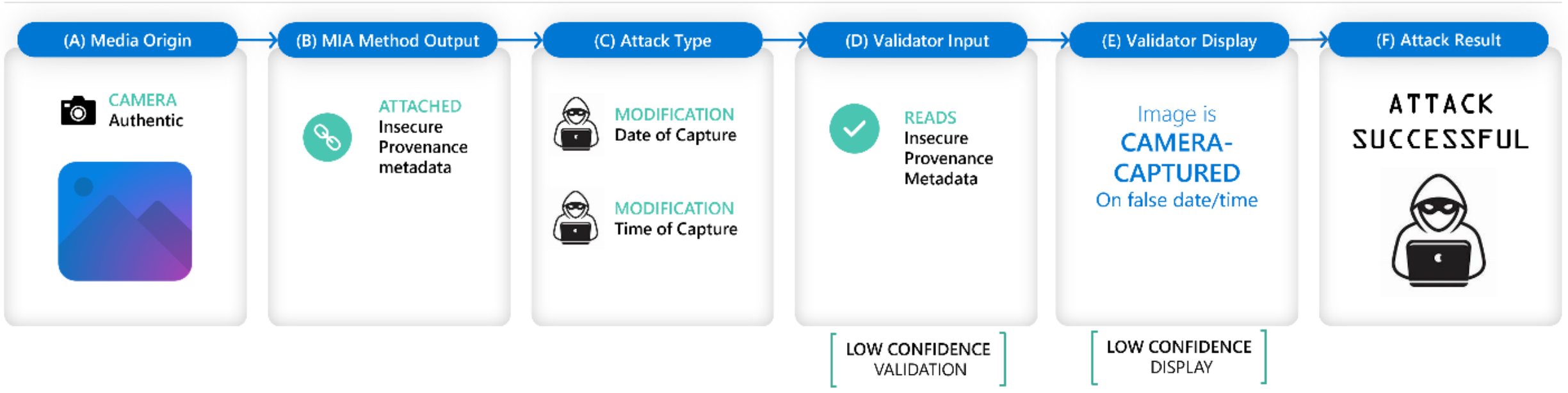

*Figure 13: The low confidence validation experience when an attacker manipulates metadata.*

## Directions to mitigate sociotechnical attacks

To mitigate potential public confusion and erosion of trust in provenance resulting from the attacks we explored, we offer the following directions:

1. **Region of Interest** ▸ Verification site providers should consider displaying details about *where* edits occur within the media, and when possible, thumbnails of media inputs, to help users, including those performing forensics and fact-checking, to interrogate the manifest and determine for themselves the extent to which such edits were significant or might affect the meaning of the media.

   There is a widely acknowledged need to differentiate editorial edits (e.g., minor touch-ups[27]) from non-editorial, material edits (e.g., removing a person or swapping a face) – as well as an awareness of the complexities of doing so in practice.[28] As the media integrity community grapples with this challenge, conveying edits made and where (referred to the C2PA as 'region of interest') via user interfaces will help parties weigh the significance of those edits. In addition to displaying such information, we recommend that verification site(s) also provide explanatory information on scenarios where they are unable to conclusively determine a validation result with high confidence.

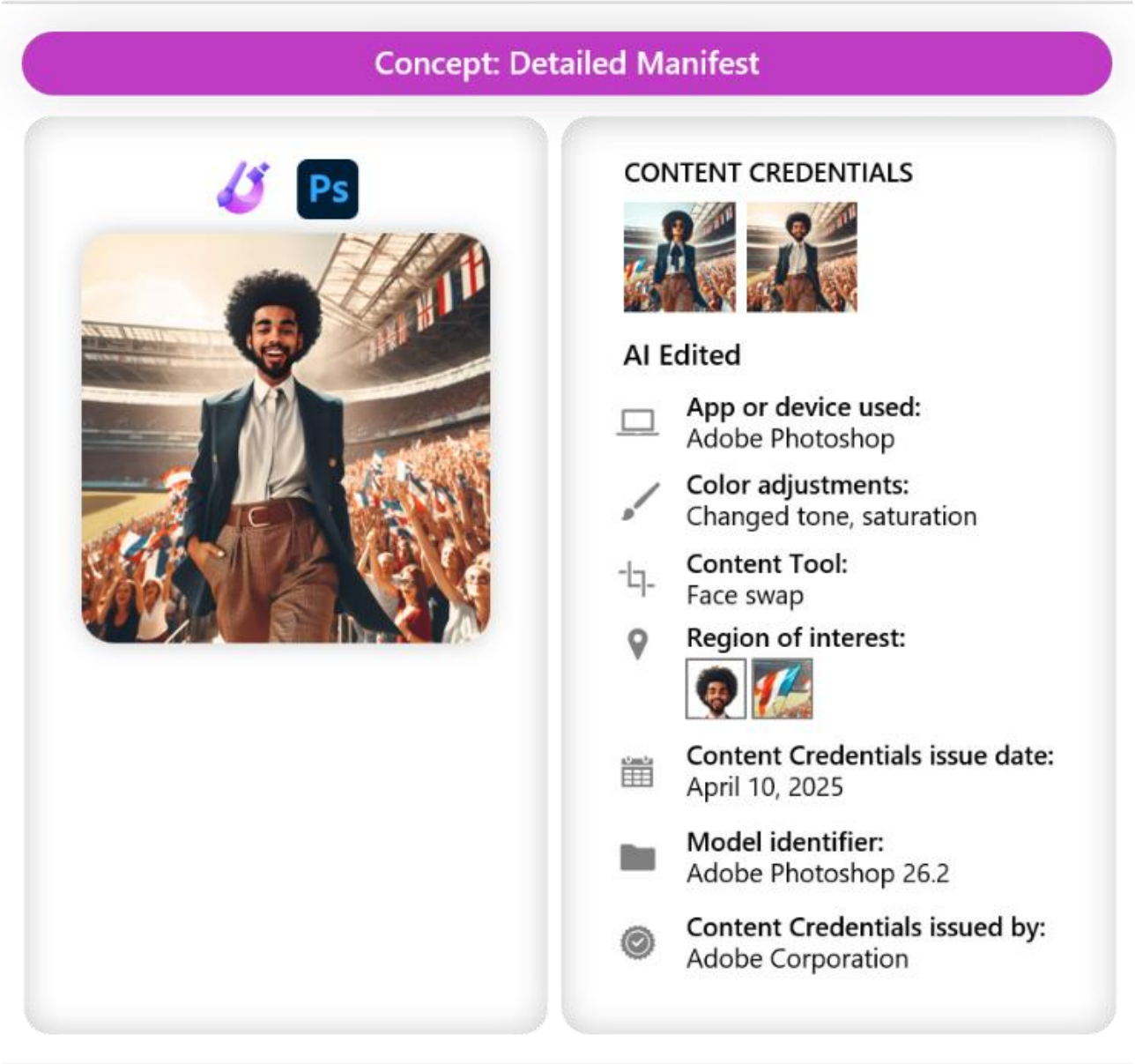

*Figure 9: A concept design that imagines a more detailed manifest with additional context.*

[27] Such touch-ups may be automatically applied by AI-enabled photo applications without a user realizing AI was used in the process.
[28] See Claire Leibowicz and Christian Cardona. Towards Responsible AI Content, November 2024. https://partnershiponai.org/resource/policy-recommendations-from-5-cases-implementing-pais-synthetic-media-framework/

2. **Manifest Preservation** ‣ Distribution platforms (i.e., social media sites) should preserve details about where edits were made to media by enabling users to download complete manifest details or explore them via other tools.

   While displaying full manifest details may not be practical in all cases, such as user feeds, this would enable users to run the image through a verification site for helpful forensics-level details. Such platforms should also explore adding C2PA manifests in cases where media is edited or transformed after uploading.

3. **UX Design** ‣ C2PA should push for research-based UX standards for consistent and effective provenance display across platforms, with iterative assessment on whether provenance display is addressing user needs. This is especially critical given the vastly different approaches platforms have taken to date (see Appendix 5), making navigating and understanding provenance information challenging for content consumers.

   At the same time, regulators requiring perceptible markings should support the adoption of a standardized mark that is designed for consistent interpretation globally and to mitigate confusion when such marks are inevitably attacked. Future directions to mitigate such attacks include platforms providing perceptible indicators of provenance for media uploads based on the authenticated, secure provenance information they contain, and algorithmic verification to assess if a pre-applied perceptible watermark on an asset aligns with the asset's authenticated, high-confidence provenance information.

4. **Security** ‣ C2PA must ensure that signing certificates accurately represent the security a hardware device or software application truly offers. Trusted signer lists that validation sites depend on must be updated regularly based on incident remediation.

   To reinforce trust in provenance validation, the C2PA Conformance Program (launched in 2025) certifies generators, validators, and certificate authorities (CAs) against the C2PA specification. Only certificates and products listed on the official C2PA trust list are recognized as conformant, to help ensure that provenance signals are technically robust and governed by a transparent, industry-wide assurance process. The program also governs certificate issuance, revocation, and periodic trust list updates, providing a foundation for high-confidence validation across the ecosystem.

5. **State-of-the-Art (SOTA) Implementations** ‣ Cameras should use secure metadata (e.g., secure implementations of C2PA-based provenance) to mitigate manipulated provenance information being displayed to content consumers. Online platforms consuming and relaying provenance information should, in turn, explore ways to differentiate between secure and insecure provenance information.

## Mitigating Additional Sociotechnical Risks

While this report focuses on how to best leverage media integrity and authentication methods to support provenance disclosure (to deliver reliable results to the public) and traceability (e.g., for forensics efforts), a workstream of this study also explored the potential impact of these technologies, once adopted, in mitigating societal and corporate risks.[29] Potential impacts, which we expect would vary across the technologies by risk area, include a deterrent effect (discouraging and preventing attackers from generating the content and causing harm because of consequences) and potential benefits for content moderation (helping platforms prevent the distribution of problematic content on social media and other platforms through blocking and moderation). An internal analysis showed varying expected results based on the different media authentication methods being employed in isolation, and that the use of multiple media integrity and authentication technologies potentially provided additive benefits. For some risk categories, the expected benefits for mitigating downstream harm were low for each MIA method.

Beyond the central focus of this report, on technical mechanisms and advances, it will be important to continue to invest in psychological and social studies and defenses aimed more centrally at exploring the understandings, skepticism, and investigative pursuits by end users. Work includes efforts to understand and iterate on information and designs for signaling provenance. Continued efforts are also needed in media and AI literacy and education.

# VI. The Limitations of Local Provenance Implementations

## Limitations for Secure Local C2PA Signing and Validation

High-confidence authentication results are possible when synthetic media has been created, provenance information added, and validation is performed in a high-security environment. Applications and services in data centers enjoy very high levels of protection for the "claim generators" that assemble C2PA manifests and for the cryptographic keys that sign them. There are many layers of security, but the most important is that the service administrators (cloud infrastructure administrators and the administrators responsible for maintaining the C2PA signing service) are motivated and can be trusted to ensure that the service operates safely and properly.

In contrast, local implementations (whereby media is generated, provenance information is added, and validation occurs offline on the client) are generally the least secure. Most (non-mobile) edge devices are administered by the owner of the device, and most current operating systems grant unlimited power to the local administrator; this includes privileges to replace/modify/debug any program[30] running on the system, including a C2PA claim generator. Similarly, although most platforms have hardware protection for cryptographic keys that prevent the key from being exfiltrated, available protections to stop a key being used by an unauthorized application are very limited. Such key use by an unauthorized application (under the control of an attacker) would enable the application to sign on behalf of the real user or bypass security guardrails (e.g., having the application generate something it would not otherwise).

---

[29] 14 risk categories were prioritized, grounded in known examples and incidents and research on relevant topics such as mis/disinformation. These risk areas include: harassment, defamation and reputation destruction, blackmailing and extortion, reputation damage to corporate and brands, IP risks, non-consensual intimate imagery (NCII), child sexual abuse material (CSAM), disinformation, tactical incitement of violence or fear, liar's dividend, psychological harm, fairness-related issues and discrepancies in the effectiveness of media integrity technologies, targeted content tailored to a specific individual or community, and phishing scams and campaigns.

Thus, in cases where a C2PA manifest is present and the manifest is validated (i.e., the signature validates as the signer is on a trust list, and hard hash contained in manifest matched the hash computed on the content), we can determine that the media is unchanged since the creation of the manifest. However, the assertions in the manifest are not necessarily accurate. (See Figure 15.) We find that the reliability of manifest information is platform-dependent with large potential variance across the ecosystem.

Given the above, enabling C2PA manifest signing on the edge can be expected to result in content with misleading or inaccurate provenance. This will include AI content with no manifest and non-AI content that is marked as synthetic. This stands the risk of diluting and undermining the security reputation of cloud-hosted provenance work. C2PA is defining levels of security to disambiguate this scenario, but some of these levels are still forthcoming, and their display and levels of understanding by the general public are not yet known.

Validation is likewise at risk, with the ability for a malicious local actor to intercept and change results from a validation process. Doing so may result in an insecure communication channel being exploited to return an inaccurate result from a properly operating validation system. Further, attempts to validate content must take into account the robustness level at which the manifest was initially generated, lest the result given to a user appear to be more definitive than warranted. Because these issues are also platform- and design-dependent, significant care and consideration of threats must be applied to designing edge scenarios.

Further, all media integrity technologies come with the inherent limitations of secure computing on the edge.

---

[30] For instance, even trusted Internet browsers (e.g. Edge and Chrome) can be hacked by malware to bypass security/provenance checks. See:
Anthony Spadafora. Chrome and Edge users infected with malicious browser extensions that steal your personal data — what to do now, August 2024. https://www.tomsguide.com/computing/malware-adware/chrome-and-edge-users-infected-with-malicious-browser-extensions-that-steal-your-personal-data-what-to-do-now?form=MG0AV3.

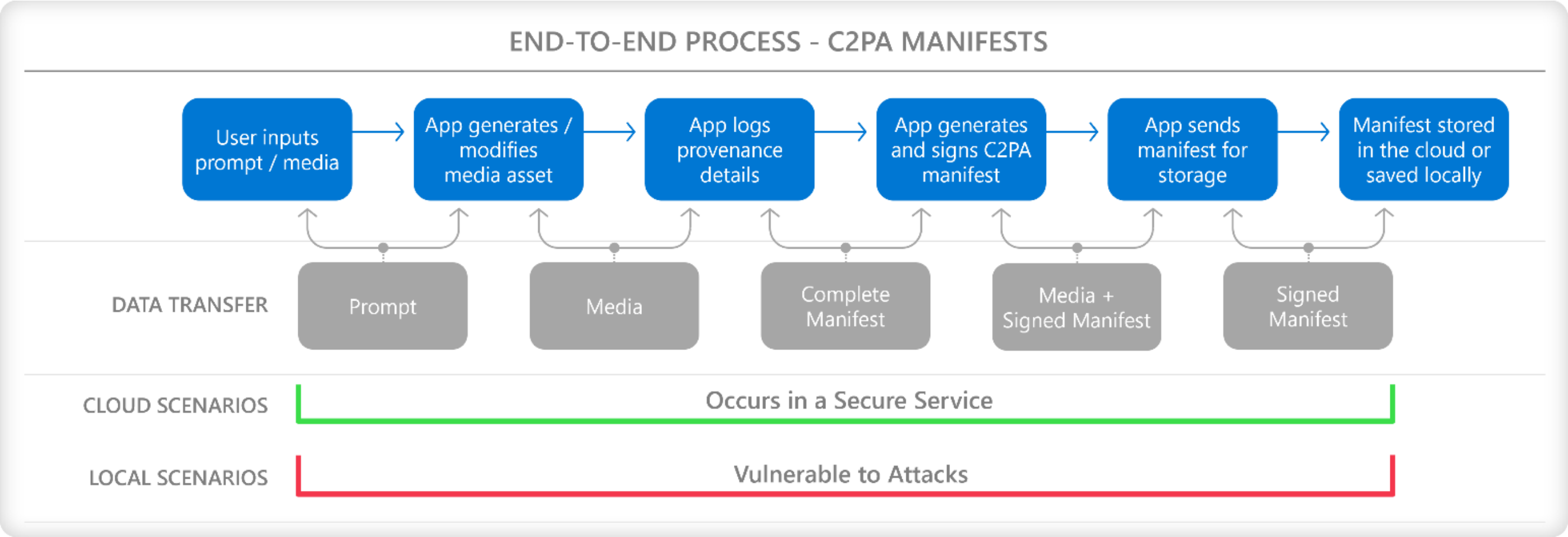

*Figure 10: Visual representation of the end-to-end process (blue) for creating, signing and storing C2PA manifests for GenAI media. Depending on platform implementation, communication channels (gray) may be secure (ex: cloud scenarios) or less secure (ex: local scenarios), depending on platform implementations. The data transfer elements may be at risk of manipulation if the communication channels are not secured.*

There are technologies that are designed to improve the security posture of edge-programs running on general purpose operating systems. These technologies serve as the building blocks for what it would take to do the 'best we can' for local implementations and still come with caveats and security limitations. Together, they would serve as secure system enablers for 1P apps and select 3P apps.

- The first is *secure certificate storage*. Client-side signing requires client-side storage of a signing certificate to prevent the signing key from being extracted. Solutions for this, such as storage with the Trusted Platform Module (TPM) or other secure enclaves, exist today and could be leveraged, along with TPM-backed keys.
- The second is *secure claim generation and insertion*. This component would create the actual manifest, with assertions both passed in from the caller and gathered by the machine as needed. It would then sign the manifest, embed it into the content, and return the resulting media. Technologies to create this component at various levels are available today through secure containers, trusted execution engine technologies, and the like.

These two building blocks provide a relatively secure method of creating, signing, and inserting a claim. They can be built using existing technology at kernel-level or higher security levels today. However, a general-purpose system for secure signing must account for the inputs into the secure enclave. This raises a host of issues, including protection of the communication path with the secure component and protection of the application itself.

The problem is relatively tractable on current mobile devices[31]: both Android and iOS can distinguish rooted from non-rooted devices and provide an execution environment that is well protected from other applications and the owner of the platform. These devices also provide attestation-style services that can cryptographically report whether the expected Claim Generator is running without modification or interference.

The problem of signing on the edge is especially hard for PC-style devices, as it is difficult to secure code and data in devices in which the end user has administrator access. Doing so would require significant re-architecture of most applications, up to and including running portions of the application in secure enclaves. It may also include requiring secure input channels form various sensors, limiting the applications able to call the secure components to a known list, barring user-generated assertions, and the like.

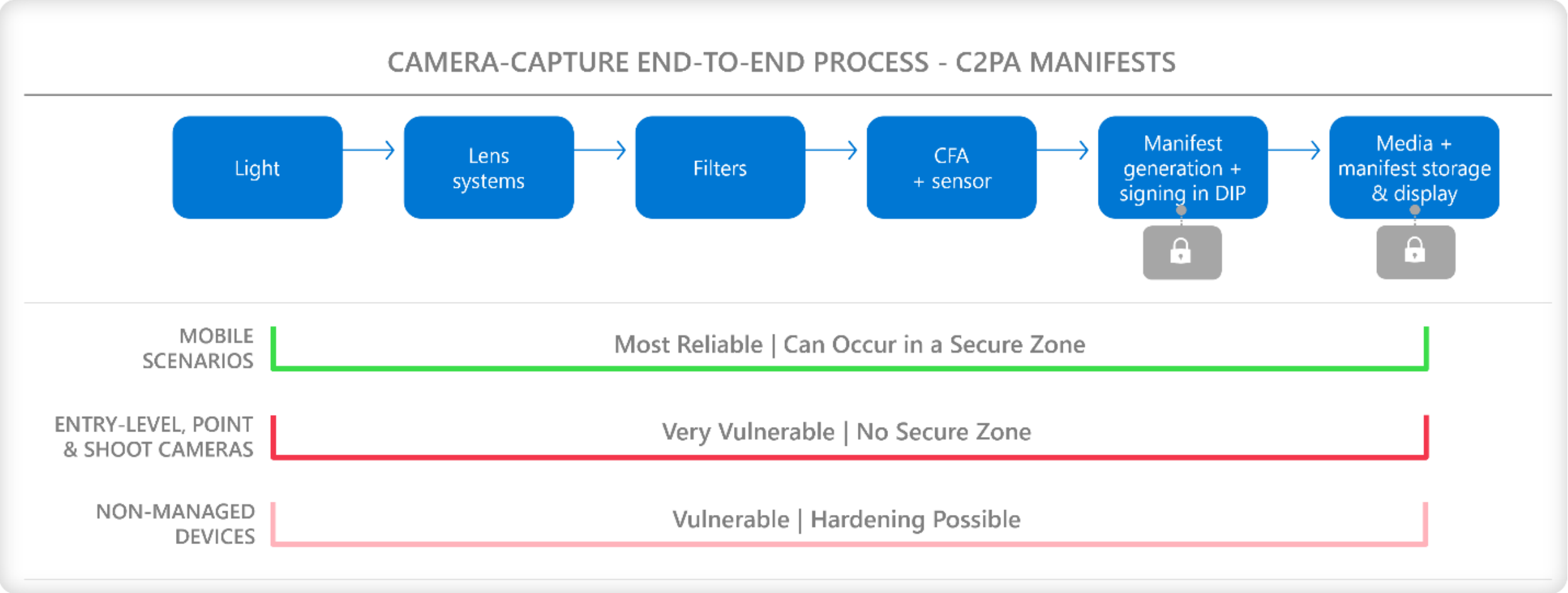

*Figure16 11: An illustration of the end-to-end process during which manifests are created, signed and stored for camera-captured media and how device variations impact the security of this provenance information. We differentiate low-end, point and shoot cameras from high-end cameras (e.g., used by news media) which may include secure chips/protected zones.*

Verification of signed data on edge devices is less problematic. A secure C2PA validator, similar to the claim generator above, could be created with existing technology. Generalized application access to that functionality is not a security issue because owners and users would only "be attacking themselves." Attack scenarios still remain in this case (for example, intercepting the SDK call for validation and changing the result). However, the OS- or firmware-provided components would reach a known security level, and statements about their reliability could be made.

## Additional considerations for watermarking and fingerprinting

Secure containers/enclaves would also contribute to more secure watermarking. Fingerprinting could also be run on a secure enclave which would limit exposure of the fingerprinting algorithm.

[31] See, for example, Eric Lynch and Sherif Hanna. How Pixel and Android are bringing a new level of trust to your images with C2PA Content Credentials, September 2025. https://security.googleblog.com/2025/09/pixel-android-trusted-images-c2pa-content-credentials.html

However, database storage of the fingerprints (soft hashes) and validation of the hashes would all still need to occur in the cloud and most scenarios demand widely available detectors, which limits the security value of protecting the fingerprint insertion algorithm and keys.

## Important use cases for local provenance

Access to the most secure, online provenance services will not be feasible nor desirable in all scenarios, including many high-risk situations. This creates a rising need for secure provenance implementations for local capture (e.g., of authentic media) to be broadly accessible, including in areas with intermittent or interrupted internet access. Increased access to such device implementations, alongside complementary connectivity technologies, will be important for ensuring that trusted provenance information can be captured and shared.

## Directions to enable more trusted provenance on edge devices

1. **Disclosure in Low-Security Environments** ▸ Device providers should explore using version 2.3 or a later version of the C2PA specification, which allows implementers to obtain signing certificates that reflect the security state of the environment for manifest generation and signing that occurs offline. This will be important to mitigate dilution of the provenance ecosystem with low-confidence results.

2. **Display in Low-Security Environments** ▸ Verification tools should show validated provenance information derived from offline devices for the highest confidence validation pathways (i.e., C2PA manifest validation, or watermark verification to recover a valid C2PA manifest). Validation tool providers should also explore displaying provenance information in a way that mitigates overreliance if the provenance was signed with a low security level.

   UI disclaimers or other measures could help mitigate overreliance; such measures should be informed by UX research to ensure they bring meaningful benefit to end users.

3. **Conformance and Display Alignment** ▸ As noted in Section V, directions 3 and 4, the C2PA should align on research-informed practices for the future of provenance display and C2PA conformance requirements. As part of this effort, the C2PA should carefully shape how security levels for provenance signing certificates impact provenance display, as guidance for this does not yet exist but will be critical for avoiding misplaced trust and overreliance in provenance results.[27]

   As part of its conformance program, C2PA is defining security levels that provenance signing certificates will have (e.g., based on product/application-level security) and UI guidance for how C2PA information should be displayed to content consumers. Stakeholders should remain heavily involved in C2PA decisions on the above to ensure certificate-level security decisions reflect diverse application and signing scenarios (e.g., cloud, hybrid, and offline) and to help ensure user confidence in provenance information they see is appropriately calibrated. Every distribution platform may choose a different way to handle security-level information (e.g., displaying it in the UI or choosing not to display manifests with low security levels) with inconsistencies likely resulting in confusion by content consumers.

# VII. AI-Based Detectors: Complementary Role and Concerns

## The Complementary Role of Detectors

While the recommendations in this report support higher confidence provenance, including the stacking and linking of technologies for added robustness and recoverability, we can expect that sophisticated actors (including nation state and organized crime actors) will be able to remove and/or undermine all media integrity and authentication methods. Further, malicious content generated by open-source models will not carry these disclosures. Thus, AI detection tools[32] can play a role as an additional line of defense when seeking to identify if content is synthetic, what model may have been used, and potentially refute that a 1P model/system was used. AI detection tools are, and will continue to be, an important tool for forensics experts who know how to interrogate detector results and are familiar with how these tools they can fail.

Microsoft work, led by the company's *AI for Good* team, has found that proprietary detectors can be valuable complements to provenance technologies for both images and videos.[33] Based on the team's analyses to date, we can speculate that, for known generators in a non-adversarial scenario (i.e., the media was not manipulated to fool detectors), accuracy could be in the ballpark of +95%.[34] In contrast, off-the-shelf AI-generated image detectors have been found to have significantly lower performance.[35]

## Challenges and Concerns with Detectors

*However, detectors come with serious limitations*. For one, because AI generators and detectors will always be in a continual "cat-and-mouse" race, we cannot rely on detectors for high-confidence assurance; they cannot be 100% reliable. The behavior of detectors is characterized by two key types of failures: false-negative rates (failures to detect synthetic or manipulated content) and false-positive rates (labeling authentic content as synthetic or manipulated).

---

[32] In using the term AI detection tools, we are referring to algorithms built to detect generative AI media in general, not algorithms built to detect a specific signature or hidden watermark intentionally embedded in a given media asset.

[33] Fake audio detection is known to be a difficult problem due to high variability in the audio outputs that need to be covered: e.g., languages, accents, tones of voices (young, old, women, men), quality of recording, and files compression. For audio, biometrics authentication and anti-spoofing technologies (that assess if a voice matches a pre-recorded sample, if there are traces of manipulation or voice cloning) can be helpful when coupled with strong authentication (e.g., multi-factor authentication). For more on the limitations of audio deepfake detection, see Menglu Li, Yasaman Ahmadiadli, and Xiao-Ping Zhang. A Survey on Speech Deepfake Detection. ACM Comput. Surv. 57, 7, Article 165, July 2025. https://doi.org/10.1145/3714458

[34] Certain types of GenAI media including image inpaintings and outpaintings are harder to detect.

[35] Research assessing off-the-shelf, AI-generated image detectors has found their precision, when using leading GenAI tools and challenging benchmark datasets, to be below 70%. See Shilin Yan, Ouxiang Li, Jiayin Cai, Yanbin Hao, Xiaolong Jiang, Yao Hu, and Weidi Xi. A Sanity Check for AI-Generated Image Detection. In *ICLR*, 2025. https://arxiv.org/pdf/2406.19435

*Thus, reliance on AI-based detectors brings to life a concerning and important paradox: the better the detectors perform, the more confidence there will be in their output. Yet the failures — particularly false negatives — from the highest-confidence detectors are likely to be the most trusted and, therefore, the most devastating.* Additional limitations for detectors include the need to continue to update the detectors amidst the continual arms-race with attackers and the ease with which detectors can be tricked if the detectors or their detection strategies are publicly known or reverse-engineered [36].

Potential attacks on detectors include adversarial training attacks (i.e., machine learning approaches that learn a detector's weaknesses from observing its predictions and cause it to produce erroneous output on images of interest) and sociotechnical attacks (e.g., taking a real image, inputting it into a GenAI system to get similar synthetic output, and having the output accurately identified as synthetic to dispute the authenticity of the original file). Thus, it is important to "red team" detectors and protect them *ex-ante*[37] through both technical and procedural safeguards, and to keep in mind that the only trustable protection against claims that a real image or media file is fake is the use of a digital signature per provenance tools on the original image or media file.

While AI detection will never be 100% reliable, it can have value in certain scenarios when used alongside other information-integrity tools and media forensics efforts: as a last resort when media has evaded processes for provenance signature, watermarking or fingerprinting, and as the main line of defense for adversarial use of open-source or bespoke models.

---

[36] Research has shown sophisticated attacks can render some detectors unusable, often driving the percentage of deepfakes detected (with a low false positive error setting) to below 70%, and in some cases below 30%. See Marija Ivanovska and Vitomir Štruc. On the vulnerability of deepfake detectors to attacks generated by denoising diffusion models. In *Proceedings of the IEEE/CVF winter conference on applications of computer vision*, pp. 1051-1060. 2024. https://doi.org/10.48550/arXiv.2307.05397.
For more on critical vulnerabilities, novel attacks, and the need to continually adapt adversarial defenses see: Umur Aybars Ciftci, Nicholas Solar, Emily Greene, Sophie Riley Saremsky, and Ilke Demir. Adversarial Reality for Evading Deepfake Image Detectors. In *Proceedings of the IEEE/CVF International Conference on Computer Vision (ICCV) Workshops*, 2025, pp. 1607-1618. https://openaccess.thecvf.com/content/ICCV2025W/APAI/html/Ciftci_Adversarial_Reality_for_Evading_Deepfake_Image_Detectors_ICCVW_2025_paper.html and Maryam Abbasi, Paulo Váz, José Silva, and Pedro Martins. Comprehensive Evaluation of Deepfake Detection Models: Accuracy, Generalization, and Resilience to Adversarial Attacks. Appl. Sci. 2025, 15, 1225. https://doi.org/10.3390/app15031225
[37] In experiments, we find the black-box attack success rate drops from 41% to 2% upon protection, whereby the model is trained using the adversarial images. We believe that proactive protection via red teaming strategies that simulate attacks and subsequent adversarial training of the detector will help prevent less sophisticated attacks.

**We recommend the following to improve the utility of detectors, and to help secure them, so they can be relied upon when needed:**

- We recommend careful examination of the implications of unavoidable failures to detect a portion of synthetically generated or manipulated content, particularly when confidence in a detector's performance is high.
- To improve detectors' reliability and transparency, we recommend investing in establishing robust and dynamic benchmarks.[38] Such benchmarks should help assess how well detectors perform globally, in real-world conditions and high-stakes contexts.[39]Future investments in detectors, much like MIA technologies, should be aimed at ensuring they perform as well as possible across the media transformation pipeline, within the workflow of key users, and in adversarial contexts.
- To help establish appropriate reliance in detection, we also recommend exploring further explainability of models' predictions. Explainability will remain a challenge and will become even more critical when most media will contain some AI-based alterations or improvements facilitated by smartphone and post-processing software. Making the distinction between cosmetic alterations and editorial alterations will be a difficult problem in the near future.[40]
- To safely provide access to detection capabilities—when needed—we suggest providing trusted partners with rate-limited APIs that do not disclose confidence scores, to avoid detector-in-the loop attacks. (An API that shares the detector's confidence scores with users provides substantially more information towards estimating the detector's model and approximating its' behavior, as compared to simply a "fake" or "real" label. Rate-limited APIs associated with restricted access to known partners/customers, to avoid adversaries using several accounts to overcome API rate limit, is the safest way to make use of detection tools.)[41]

---

[38] See Thomas Roca, et al. Introducing The MNW Benchmark For AI Forensics, July 2025. https://www.semanticscholar.org/paper/I-NTRODUCING-THE-MNW-B-ENCHMARK-FOR-AI-F-ORENSICS-Roca-Postiglione/0b3569e567d57b5d2692273ecefc96611d5c4cfa

[39] For examples of challenges in detector performance in real-world contexts and areas for improvement, see shirin anlen. Five Real-World Failures Expose Need for Effective Detection of AI-Generated Media, June 2025. https://www.techpolicy.press/five-real-world-failures-expose-need-for-effective-detection-of-ai-generated-media and WITNESS. New Global Benchmark for AI Detection, 2025. https://www.witness.org/ai-detection-global-benchmark-witness-2/.

[40] See Thomas Roca, Pengce Wang, Keri Mallari, Kevin White, and Juan M. Lavista Ferres. Deepfake Detection: Don't Take Video at Face Value, Or Should You? Microsoft Journal of Applied Research, Vol 22, October 2025.

[41] See Thomas Roca, Pengce Wang, Meghana Kshirsagar, and Juan Lavista Ferres. Can AI Detectors Be Protected Against Perturbation Attacks? Lessons Learned from Playing The 'Cat and Mouse' Game. Microsoft Journal of Applied Research, Vol 22, October 2025.

# VIII. Ongoing Research and Policy Development

## Directions for research investments and iterative policy efforts

For authenticity efforts to be successful, industry standards, and cross-company collaborations on them, must exist. The ecosystem of media distribution is complicated with many players, so without interoperable standards the likelihood of broad success and real benefit to users is slim. As the leading and growing standards body on provenance, and with a specification that integrates provenance, watermarking, and fingerprinting, C2PA remains an essential body for engagement on standards. We recommend the following research directions as important areas for the ecosystem to explore to improve the reliability and efficacy of provenance information.

1. **Use and Display Research** ▸ Current display is inconsistent.[42] As such, C2PA or its members should champion research workstreams to better understand the use and display of provenance signals both in the short- and long-term, and share these results with the community to improve consistency and effectiveness.
   Further research would be especially valuable in the following areas:
   - Further UX/UI research to determine how to foster appropriate trust, reliance, and understanding of provenance information displayed, including across geographic and product-specific contexts. Insights from such studies should be used to inform provenance display UI for products, C2PA UX guidance, and media literacy efforts.
   - Research on how users comprehend and respond to a mix of provenance-enabled and non-enabled content.
   - Research on how to communicate heterogeneous sets of changes to users, including mixtures of authentic and synthetic content in the same material across modalities.
   - Research to advance in-stream tools that display provenance information where people are and distinguish between high- and lower-confidence provenance signals.
2. **Manifest Stores** ▸ Further research is needed to define best practices for implementing manifest stores, including exploring a potential centralized collection of stores from multiple entities or a decentralized version. This should include use cases for closed vs. open scenarios as well as deployment best practices that account for the security and access requirements and diverse needs of those implementing the C2PA standard.
   A collection of manifest stores with access for trusted parties – or a decentralized manifest store - could potentially help mitigate sociotechnical attacks explored in Section V of this report. For example, an important open question is how to best verify if detected provenance information relates to expected provenance information. Various levels of checks could be performed by platforms to see if there is a mismatch between the manifest ingested and what exists in an external system. A higher-level matching of manifests could support the validation process - enabling checks to see if the provenance detected matches the provenance expected per proof of time and publication verification.

[42] See Appendix 5.

Adoption of C2PA by photographers could help mitigate attacks whereby camera-captured media are signed by another party as synthetic if, in the future, a distributed system exists for parties to store their manifests. Such a database could allow recovery of the original manifest signed with the date/time of singing.

3. **Continuous Feedback Cycle** ▶ The C2PA Steering Committee should review feedback from other members, researchers, civil society organizations, and the public to continue improving the standard.
   To ensure interoperability and maintain trust, stakeholders should actively engage with the C2PA Conformance Program and leverage its Conformance Explorer to verify the status of generators, validators, and CAs. This alignment is critical for scaling adoption and for ensuring that provenance signals remain credible as the ecosystem evolves.

4. **Red-Teaming and Analysis to Identify and Mitigate Weaknesses** ▶ MIA stakeholders should engage in ongoing technical and sociotechnical red-teaming and analysis to probe for weaknesses in the methods, to support transparent disclosure of strengths and weaknesses, and to guide refinements of technical approaches, policies, and laws.

   To support this, the C2PA should promote ongoing intensive red-teaming and analysis of its specifications and implementations, with an eye towards mitigating potential harms and disproportionate risks to vulnerable groups globally. Continued input from the C2PA Threats and Human Rights Task Force will be important for advancing the standard's guiding principles of respecting privacy, meeting the needs of global audiences, and mitigating potential abuse and misuse.

5. **Iterative Policy Development** ▶ Policy efforts should drive adoption of technical methods for which there is implementation readiness, while building an understanding of limitations that may exist to inform the public's interpretation of provenance reliability. Policy expectations should be incrementally lifted in tandem with advancements in research and technical methods that can be deployed at scale.

6. **Policy Accommodations** ▶ The report findings underscore the value of robust media integrity and authentication practices yet also reflect the reality that technical and operational contexts can vary widely. As the ecosystem evolves, it may be prudent for policy approaches to accommodate a range of implementation scenarios, ensuring that efforts to strengthen media authenticity remain effective and relevant across diverse environments.

## Acknowledgments

We greatly appreciate the thoughtful feedback and insightful consultations provided by Ginny Badanes, Josh Benaloh, Sarah Bird, Alon Brown, Jacobo Castellanos, Amanda Craig, Natasha Crampton, Madeleine Daepp, Chanpreet Dhanjal, Karen Easterbrook, Ani Gevorkian, Sam Gregory, Ashish Jaiman, Jay Li, Sarah McGee, Phillip Misner, Robert Ness, Nazmus Sakib, and Ray Xu.

# Appendices

**Appendix 1**

## Glossary of Terms

We generally follow glossaries of digital content transparency methods published by the National Institute of Standards and Technology (NIST) and the Partnership on AI (PAI).[43] In light of the report's exploratory scope, we've supplemented existing lists with the terms below.

**Asset**: a digital file.

**Authentication (of media)**: the process of verifying the origin, integrity, and authenticity of digital media content. Its goals are to ensure that: the content comes from a legitimate source; the content has not been altered in unauthorized ways since its creation or signing; and the identity of the signer can be cryptographically verified.

**Authentic media**: media captured by a *capture device* that includes representations of real-life scenes, people, places, and objects.

**C2PA Conformance Program:** The formal certification program operated by the Coalition for Content Provenance and Authenticity (C2PA), which verifies that generators, validators, and certificate authorities conform to the C2PA specification. The program maintains a public trust list and defines assurance levels for provenance signing certificates.

**Capture Device**: a device that can record photographs, audio, or video content such as point and shoot cameras, video cameras, or mobile phones with built-in cameras or microphones.

**Client:** a device or application operating without internet connectivity. See also *edge, local*.

**Coalition for Content Provenance and Authenticity (C2PA)**: a standards body focused on the development of open, global technical standards and specifications for establishing content provenance and authenticity.

**Content Credentials (Cr)**: an open technical standard, provided by the C2PA, for publishers, creators and consumers to establish the origin and edits of digital content. The Cr icon is used to indicate content that has been cryptographically signed using C2PA Content Credentials.

**Edge**: a device or application operating without internet connectivity. See also *client*, *local*.

**Fingerprinting (hard hashing)**: computing an identifier (i.e., hash) that is associated with an asset but created and stored outside of it. Hard hashing is used to identify an exact match. For more details, see page 11.

[43] See Bilva Chandra, Jesse Dunietz, Kathleen Roberts, Yooyoung Lee, Peter Fontana, and George Awad. Reducing risks posed by synthetic content an overview of technical approaches to digital content transparency, National Institute of Standards and Technology, 2024. https://doi.org/10.6028/NIST.AI.100-4 and Partnership on AI. Building a Glossary for Synthetic Media Transparency Methods, 2023. https://partnershiponai.org/resource/glossary-for-synthetic-media-transparency-methods-part-1/.

**Fingerprinting (soft hashing)**: computing an identifier (i.e., hash) that is associated with an asset but created and stored outside of it. Soft hashing is used to identify similar matches. For more details, see page 11.

**Local**: a device or application operating without internet connectivity. See also *client*, *edge*.

**Manifest**: the complete metadata included when an asset was signed with C2PA Content Credentials. Metadata may include assertions about the media (such as whether AI was used to generate or edit it), the app or device used (such as the camera model or AI tool used to generate all, or part, of the content), the date and time the Content Credential was created and signed, and the entity that created and signed the Content Credential and made the assertions it contains.

**Media or Digital Media or Content:** images, audio, or video.

**Metadata**: data that describes, explains, or provides context for other data. Metadata does not represent the actual content but gives details that make the content easier to organize, find, interpret, and manage. A subset of metadata relates to content's provenance.

**Non-secure Provenance:** Provenance metadata that is easy to edit or manipulate.

**Provenance**: 1. the origin and history (e.g., edits) of digital content. 2. the method of attaching metadata about the origin and history of digital content to the asset itself.

**Secure provenance**: cryptographically signed and protected metadata about the origin and history of digital content. Assertions about provenance may not be accurate but once signed by an entity, they are protected and cannot be manipulated without breaking the entity's signature. For more details on secure and non-secure provenance, see page 9.

**Sociotechnical attacks**: attacks that exploit weaknesses that emerge when technology and social behaviors intersect—for instance, exploiting trust placed in digital provenance, reliance on automated validation, or gaps in governance and user understanding.

**Synthetic media**: media that has been generated or modified via artificial intelligence.

**Watermarking**: embedding information into a digital asset that can assist with verifying its provenance. Watermarks can be perceptible or imperceptible. For more details, see page 10.

## Appendix 2: Understanding Potential Attacks on MIA Methods and Mitigations

| | CAPABILITIES | LIMITATIONS | ATTACKS *[mitigation strategy]* |
|---|---|---|---|
| **PROVENANCE** Secure Metadata | • Ideal for well-intentioned actors to disclose provenance info.<br>• Open standard enables reading and displaying provenance information at scale.<br>• Captures rich context and edits, including alterations made with generative AI tools and chain-of-custody edits made across the content's lifecycle. Exceeds info embedded in a watermark.<br>• Reliably conveys with security guarantees (a) the Content Credential was legitimately signed by the entity listed and (b) the assertions and content have not been tampered with since the signing occurred. | • Easy for general user or malicious actor to remove provenance data.<br>• Some social media sites remove metadata during upload process.<br>• Validated provenance does not prove content is true. | • Cr, including signer info, can be removed/stripped (e.g., with a screenshot, by social media platforms not ingesting it). [*Potential mitigations include editing tools and social media platforms preserving provenance manifests.]*<br>• After removal, a fake Cr can be added - e.g.: injection or manipulation of content and then re-signing. Note: the signer can't technically be forged. However, the signer can be misrepresented if a malign actor signs with a certificate that appears to come from the entity and it isn't caught/blocked by the trust root process. [Use of *C2PA v2.3 spec or a later version serves as a mitigation.]*<br>• Cr logo can be forged at the UI level (e.g., pasting the icon onto the media so Content Credentials appear to be tied to it when they are not). [*Media literacy and education about what the Cr pin does/does not indicate and how to interact with it serves as a mitigation.*]<br>• AI generated pixels can be copied/pasted to bypass inclusion of those edits in the Cr. [*Mitigated in online media generation and signing scenarios, and when tools adding C2PA manifests account for full edit history.*]<br>• False info can be added to a Cr when it is added retroactively (e.g., an entity wants to add provenance info to files created in the past). [*Media literacy serves as a mitigation if content consumers consider the signing entity and the level of trust they have in that entity.*]<br>• Specific to low-security environments: False info can be added to a Cr during the process of creating new GenAI media, with the manifest then signed and certified (e.g., by the hardware/software with a provenance signing certificate). [*Use of v2.3 or later of the C2PA spec serves as a mitigation, by differentiating manifests signed at a low-security level. Secure enclaves also serve as a mitigation.* |
| **WATERMARK** Imperceptible | • Ideal for use with provenance metadata for added robustness to potentially assist with metadata recovery if stripped.<br>• Metadata is embedded in the content itself and therefore survives most data processing pipelines.<br>• Robust, state-of-the-art watermarks can reliably be used to embed non-security critical metadata (e.g., content is AI-generated, authorship/copyright info, including a pointer to Content Credentials).<br>• Capable of withstanding more modifications/transformations than provenance metadata per C2PA. | • Detection is probabilistic so false positives and false negatives may occur.<br>• Can be forged if the watermarking algorithm is known/reverse engineered.[44]<br>• Constraints on the type and volume of data that can be embedded.<br>• Constraints on the number of times media can be watermarked across its lifecycle before watermark detection errors increase.<br>• While keeping watermarking methods private enhances security, it complicates public verification. | • Can be removed (per distortion, etc.). *[Fingerprinting as an additional cross-check as shown in Appendix 4.]*<br>• Can be forged (inaccurately adding a watermark to media in which it does not belong). *[Fingerprinting as an additional cross-check as shown in Appendix 4.]*<br>• Can be reverse engineered via decoder access (through an Oracle attack). [*Employing multiple watermarks and/or unique keys can serve as mitigations.]*<br>Note: Watermarking algorithms are easier to reverse engineer if watermarking insertion is performed on an edge device.<br>• Infrastructure attacks may also occur and should be included in the Threat Model for deployment (e.g., attacks of cloud infrastructure to get the watermarking algorithm or encoder/decoder keys). *[Having a plan to rotate to a new/different watermarking algorithm, as needed, can serve as a mitigation.]*<br>• Can be subject to software tampering when implemented on a client – e.g., to bypass watermarking insertion on the media or modify the watermarking pattern to avoid forensic tracing. *[Secure enclaves can serve as mitigation tools.]* |
| **WATERMARK** Perceptible | • Serves as an accessible way to disclose provenance information (e.g., source and synthetic nature) as it is immediately seen or heard by content consumers. | • May be viewed as detracting from the visual content on which it is overlayed.<br>• Not easily machine-readable, which can be a concern for identifying these watermarks at scale. | • Can easily be forged<br>• Can easily be removed. (NIST.AI.100-4.SyntheticContent.ipd.pdf *notes application across a large swath of the content can serve as a potential mitigation, with removal potentially corrupting the content*. *However, such extensive markings come with trade-offs for content usability.]* |

[44] Stripping the watermark pattern from one piece of media and inserting it into another is difficult to do using signal processing techniques, but not impossible, especially when combined with machine learning / AI techniques. If the forgery attack is successful, the watermark detector will detect a valid ID, but that ID belongs to another piece of media, so that provenance of the piece of media under test will be incorrectly determined.

**FINGERPRINT**
Soft Hash

- Allows for indexing and retrieval of content in a way that is independent of metadata or changes in resolution or fidelity.
- Excels at matching when non-adversarial edits occur during day-to-day handling of media files (e.g., file format conversion, minor file size size/quality reduction) and when small modifications have occurred (e.g., addition or removal of small portions of a picture).[45]
- Serves as a means of leak/traitor tracing or to look for a match to known problematic content (e.g., PhotoDNA for CSAM content; Google Content ID for copyright-protected content).
- Ability to compare versions to identify edits to the asset.

- Soft hashing is non-unique; multiple matches may be returned or errors may occur, making human verification important in high-stakes and security-critical contexts.
- Subject to 'hash collisions' where two perceptually different input files have the same hash resulting in false matches.[46]
- Cost and complexity with maintaining the requisite database.
- Potential inability to migrate a hash database when transitioning from one hash to another (e.g., phase out a hash function deemed ineffective, inefficient or insecure).
- Unable to recognize whether the same component is in two different files unless the entirety of both files is very similar (e.g., similar angels, lighting, environment, background).

- Subject to attacks on content to exploit the perceptual hash function – i.e., manipulation / precise perturbation of images and videos to prevent detection of harmful content (false negatives) or to cause the misclassification of benign content (false positives). *[Increasing the resolution of image blocks on which the hash is computed, and thus increasing the length of the soft hash, could serve as a potential mitigation, acknowledging trade-offs with higher computation cost and storage requirements.]*
- Manipulated content may be perceptually indistinguishable from the original. Such media manipulation/perturbation may occur:
  a) after the hash has been added to a database, or b) before being added to the database to create a hash collision with some other, targeted media item that does not possess the characteristics necessary for inclusion in the database. [*Human-in-the-loop verification can serve as a mitigation for hash collisions.]*
- Hash database may be attacked (e.g., to delete content from or add inappropriate content to the database). *[Proper access controls, logging, detection, and backups applied to the database can serve as mitigations.]*
- Can be subject to client-side specific attacks including:
  - Reverse engineering to extract (parts of) the hash function or hash database from the source code of the application or operating system. *[Changing out hash functions, employing code obfuscation techniques, and secure enclave execution can serve as mitigations.]*
  - Software tampering – e.g., to change the output of the soft hashing function so an incorrect hash gets sent to the cloud database. [S*ecure enclaves/zones can serve as mitigation tools.]*

---

[45] As such, they can help protect against watermark forgery.

[46] Pixel-value modifications can be made to lead to the same soft hash for two distinct pieces of media.

## Appendix 3: Disclosure Outcomes: Possible Online Cases

The table below depicts the validation results possible for a media asset if sequential validation of C2PA manifests, watermarks, and fingerprints are performed, considering the attacks and verification errors that may occur for each MIA method. While the table attempts to cover many possible error conditions and attacks, it is meant to be exemplary and should not be considered exhaustive.

As displayed in the table below, after the presence of a C2PA manifest is verified, the presence of watermark is verified, or a fingerprint match is verified, an additional step is performed to cross-check the C2PA hash. In the case of C2PA verification, we confirm that the hash computed for the media uploaded matches the hash found inside the C2PA manifest. After watermark and fingerprint verification, we confirm which C2PA manifest in our internal manifest store corresponds to the watermark ID or fingerprint hash, and then cross-check to see if the hash computed for the media uploaded matches the hash found the C2PA manifest stored on the server.

### Verification Legend

**Indeterminate**: cannot determine provenance (e.g., manifest is missing)
**Media Modified**: can't make claims about provenance/verify who produced it
**Possible Match**: may be able to make claims pending a human-in-the-loop review*
**Match**: exact copy of what entity produced
**Media Validates**: manifest is present on an exact copy of what (the signed) entity produced

** If manually verified, confidence could be higher than 'lowest confidence', but we presume an automated process for the public validation tool and state the confidence level accordingly.*

| | C2PA VERIFICATION | | WATERMARK VERIFICATION | | FINGERPRINT VERIFICATION | | FINAL VERIFICATION | CONFIDENCE | CONCERNS |
|---|---|---|---|---|---|---|---|---|---|
| **#** | **Check-1 →** | **Check-2 →** | **Check-1 →** | **Check-2 →** | **Check-1 →** | **Check-2 →** | **= Result** | **= Level** | **Potential Attacks or Errors** |
| 1 | *Manifest:* Not Present | | Detectable | *C2PA Hash:* No Match | Invalid | | Indeterminate | Low | |
| 2 | *Manifest:* Not Present | | Detectable | *C2PA Hash:* No Match | Valid | *C2PA Hash:* No Match | Media Modified | Low | Valid fingerprint but still low confidence it is the version the signing entity generated. Minor modifications may not alter the fingerprint while altering the semantic meaning. |
| 3 | *Manifest:* Not Present | | Detectable | *C2PA Hash:* No Match | Valid | *C2PA Hash:* Match | Possible Match | Low | |
| 4 | *Manifest:* Not Present | | Detectable | *C2PA Hash:* No Match | Valid | *C2PA Manifest:* Missing from Registry | Indeterminate | Low | Watermark forgery or potential timeout error |

| | | | | | | | | |
|---|---|---|---|---|---|---|---|---|
| 5 | *Manifest:* Not Present | Detectable | *C2PA Hash:* No Match | No Access | | Media Modified | Low | |
| 6 | *Manifest:* Not Present | Detectable | *C2PA Hash:* Match | Valid | *C2PA Hash:* Match | Match | High | |
| 7 | *Manifest:* Not Present | Detectable | *C2PA Hash:* Match | No Access | | Match | High | |
| 8 | *Manifest:* Not Present | Detectable | *C2PA Manifest:* Missing from Registry | Invalid | | Indeterminate | Cannot Be Asserted | |
| 9 | *Manifest:* Not Present | Detectable | *C2PA Manifest:* Missing from Registry | Valid | *C2PA Hash:* No Match | Indeterminate | Lowest | |
| 10 | *Manifest:* Not Present | Detectable | *C2PA Manifest:* Missing from Registry | Valid | *C2PA Hash:* Match | Possible Match | Lowest | Depends on granularity of fingerprints; human-in-the loop (HITL) review required |
| 11 | *Manifest:* Not Present | Detectable | *C2PA Manifest:* Missing from Registry | Valid | *C2PA Manifest:* Missing from Registry | Indeterminate | Lowest | Potential registry timeout (error) |
| 12 | *Manifest:* Not Present | Detectable | *C2PA Manifest:* Missing from Registry | No Access | | Indeterminate | Cannot Be Asserted | Probable registry timeout |
| 13 | *Manifest:* Not Present | No Access | | Invalid | | Indeterminate | Cannot Be Asserted | |
| 14 | *Manifest:* Not Present | No Access | | Valid | *C2PA Hash:* No Match | Media Modified | Lowest | |
| 15 | *Manifest:* Not Present | No Access | | Valid | *C2PA Hash:* Match | Possible Match | Lowest | Depends on granularity of fingerprints; human-in-the loop (HITL) review required |
| 16 | *Manifest:* Not Present | No Access | | Valid | *C2PA Manifest:* Missing from Registry | Indeterminate | Lowest | Probable registry timeout, watermark removal, fingerprint hash collision |
| 17 | *Manifest:* Not Present | No Access | | No Access | | Indeterminate | Cannot Be Asserted | |
| 18 | *Manifest:* Not Present | Undetectable | | Invalid | | Indeterminate | Cannot Be Asserted | Probable watermark removal; HITL review required |
| 19 | *Manifest:* Not Present | Undetectable | | Valid | *C2PA Hash:* No Match | Media Modified | Lowest | |
| 20 | *Manifest:* Not Present | Undetectable | | Valid | *C2PA Hash:* Match | Possible Match | Lowest | Depends on granularity of fingerprints; HITL review required |
| 21 | *Manifest:* Not Present | Undetectable | | Valid | *C2PA Manifest:* Missing from Registry | Indeterminate | Lowest | Probable registry timeout and watermark removal |

| | | | | | | | | | |
|---|---|---|---|---|---|---|---|---|---|
| 22 | *Manifest:* Not Present | | Undetectable | | No Access | | Indeterminate | Cannot Be Asserted | |
| 23 | Present | *C2PA Hash:* No Match | Detectable | *C2PA Hash:* No Match | Invalid | | Media Modified | Low | |
| 24 | Present | *C2PA Hash:* No Match | Detectable | *C2PA Hash:* No Match | Valid | *C2PA Hash:* No Match | Media Modified | Low | |
| 25 | Present | *C2PA Hash:* No Match | Detectable | *C2PA Hash:* No Match | Valid | *C2PA Hash:* Match | Possible Match | Low | Likely C2PA replacement attack, and/or possible watermark attack or fingerprint collision. |
| 26 | Present | *C2PA Hash:* No Match | Detectable | *C2PA Hash:* No Match | Valid | *C2PA Manifest:* Missing from Registry | Indeterminate | Low | Watermark forgery, potential C2PA replacement, probable timeout |
| 27 | Present | *C2PA Hash:* No Match | Detectable | *C2PA Hash:* No Match | No Access | | Media Modified | Low | |
| 28 | Present | *C2PA Hash:* No Match | Detectable | *C2PA Hash:* Match | Valid | *C2PA Hash:* Match | Match | High | C2PA replacement attack |
| 29 | Present | *C2PA Hash:* No Match | Detectable | *C2PA Hash:* Match | No Access | | Match | High | |
| 30 | Present | *C2PA Hash:* No Match | Detectable | *C2PA Manifest:* Missing from Registry | Invalid | | Indeterminate | Cannot Be Asserted | |
| 31 | Present | *C2PA Hash:* No Match | Detectable | *C2PA Manifest:* Missing from Registry | Valid | *C2PA Hash:* No Match | Media Modified | Lowest | |
| 32 | Present | *C2PA Hash:* No Match | Detectable | *C2PA Manifest:* Missing from Registry | Valid | *C2PA Hash:* Match | Possible Match | Lowest | Manifest cannot be both in and not in the registry. Potential metadata manifest and watermark replacement. |
| 33 | Present | *C2PA Hash:* No Match | Detectable | *C2PA Manifest:* Missing from Registry | Valid | *C2PA Manifest:* Missing from Registry | Indeterminate | Lowest | Probable registry timeout |
| 34 | Present | *C2PA Hash:* No Match | Detectable | *C2PA Manifest:* Missing from Registry | No Access | | Indeterminate | Cannot Be Asserted | Probable timeout |
| 35 | Present | *C2PA Hash:* No Match | No Access | | Invalid | | Indeterminate | Cannot Be Asserted | |
| 36 | Present | *C2PA Hash:* No Match | No Access | | Valid | *C2PA Hash:* No Match | Media Modified | Lowest | |
| 37 | Present | *C2PA Hash:* No Match | No Access | | Valid | *C2PA Hash:* Match | Possible Match | Lowest | Processing error/ error entering info in database |
| 38 | Present | *C2PA Hash:* No Match | No Access | | Valid | *C2PA Manifest:* Missing from Registry | Indeterminate | Lowest | Probable registry timeout |
| 39 | Present | *C2PA Hash:* No Match | No Access | | No Access | | Indeterminate | Cannot Be Asserted | |
| 40 | Present | *C2PA Hash:* No Match | Undetectable | | Invalid | | Indeterminate | Cannot Be Asserted | |
| 41 | Present | *C2PA Hash:* No Match | Undetectable | | Valid | *C2PA Hash:* No Match | Media Modified | Lowest | Watermark removal attack or benign modification that changed the media enough for the watermark to become undetectable. |
| 42 | Present | *C2PA Hash:* No Match | Undetectable | | Valid | *C2PA Hash:* Match | Possible Match | Lowest | |
| 43 | Present | *C2PA Hash:* No Match | Undetectable | | Valid | *C2PA Manifest:* Missing from Registry | Indeterminate | Lowest | Potential registry timeout |
| 44 | Present | *C2PA Hash:* No Match | Undetectable | | No Access | | Indeterminate | Cannot Be Asserted | |
| 45 | Present | *C2PA Hash:* Match | Detectable | *C2PA Hash:* Match | Valid | *C2PA Hash:* Match | Media Validates | High | |
| 46 | Present | *C2PA Hash:* Match | Detectable | *C2PA Hash:* Match | No Access | | Media Validates | High | |
| 47 | Present | *C2PA Hash:* Match | Detectable | *C2PA Manifest:* Missing from Registry | Invalid | | Media Validates | High | |
| 48 | Present | *C2PA Hash:* Match | Detectable | *C2PA Manifest:* Missing from Registry | Valid | *C2PA Hash:* Match | Media Validates | High | |
| 49 | Present | *C2PA Hash:* Match | Detectable | *C2PA Manifest:* Missing from Registry | Valid | *C2PA Manifest:* Missing from Registry | Media Validates | High | |
| 50 | Present | *C2PA Hash:* Match | Detectable | *C2PA Manifest:* Missing from Registry | No Access | | Media Validates | High | |
| 51 | Present | *C2PA Hash:* Match | No Access | | Invalid | | Media Validates | High | |
| 52 | Present | *C2PA Hash:* Match | No Access | | Valid | *C2PA Hash:* No Match | Media Validates | High | |
| 53 | Present | *C2PA Hash:* Match | No Access | | Valid | *C2PA Hash:* Match | Media Validates | High | |
| 54 | Present | *C2PA Hash:* Match | No Access | | Valid | *C2PA Manifest:* Missing from Registry | Media Validates | High | |
| 55 | Present | *C2PA Hash:* Match | No Access | | No Access | | Media Validates | High | |

| | | | | | | | | |
|---|---|---|---|---|---|---|---|---|
| 56 | Present | *C2PA Hash:* Match | Undetectable | Invalid | | Media Validates | High | |
| 57 | Present | *C2PA Hash:* Match | Undetectable | Valid | *C2PA Hash:* No Match | Media Validates | High | |
| 58 | Present | *C2PA Hash:* Match | Undetectable | Valid | *C2PA Hash:* Match | Media Validates | High | |
| 59 | Present | *C2PA Hash:* Match | Undetectable | Valid | *C2PA Manifest:* Missing from Registry | Media Validates | High | |
| 60 | Present | *C2PA Hash:* Match | Undetectable | No Access | | Media Validates | High | |

## Appendix 4: Holistic View of Potential Validation Results

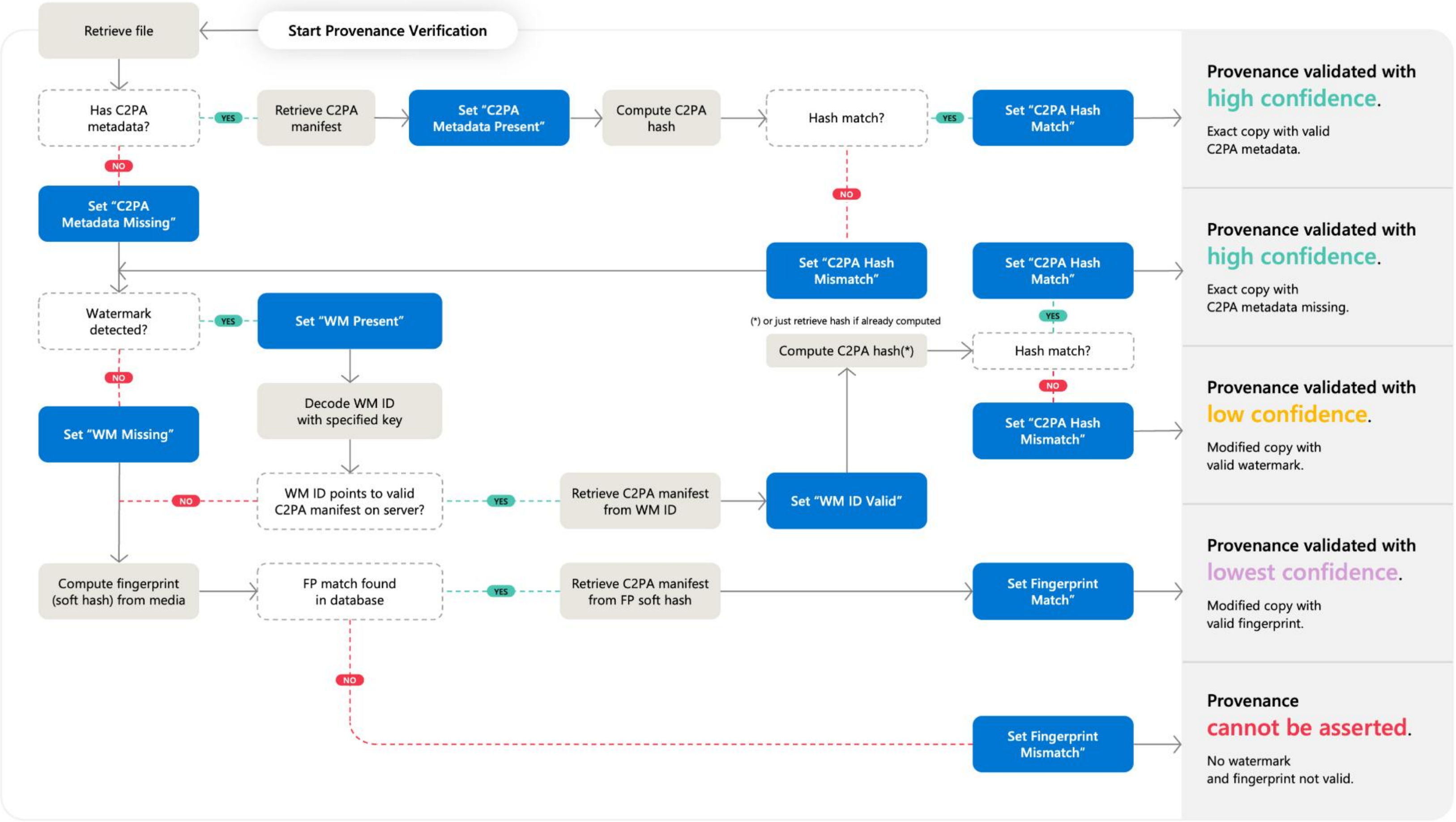

## Appendix 5: Varied Provenance "Signal" Display Across Distribution Platforms

The examples below show how provenance display information, both in terms of granularity and prominence, varies across platforms. We expect provenance display will be an area of critical innovation. At the same time, we expect UX considerations will become more complicated and display implementations more varied as (1) modification history becomes more complex to covey, (2) security levels for provenance are introduced, and (3) platforms weigh buried vs. prominent display based on decisions about what content is sensitive and what edits are material. The image shared across platforms was created using Microsoft Designer with Content Credentials applied on February 17, 2026.

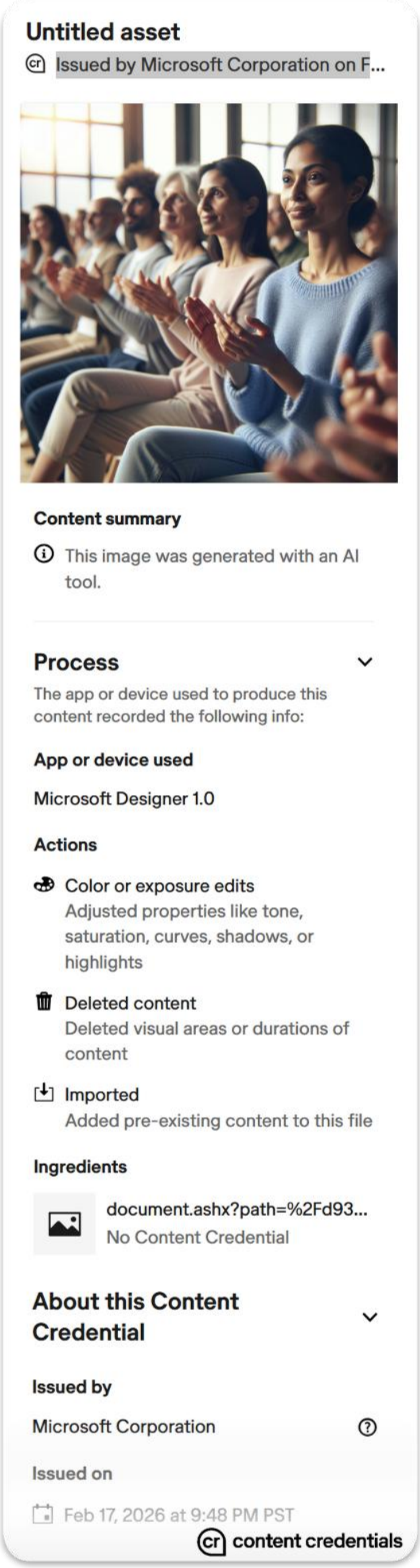

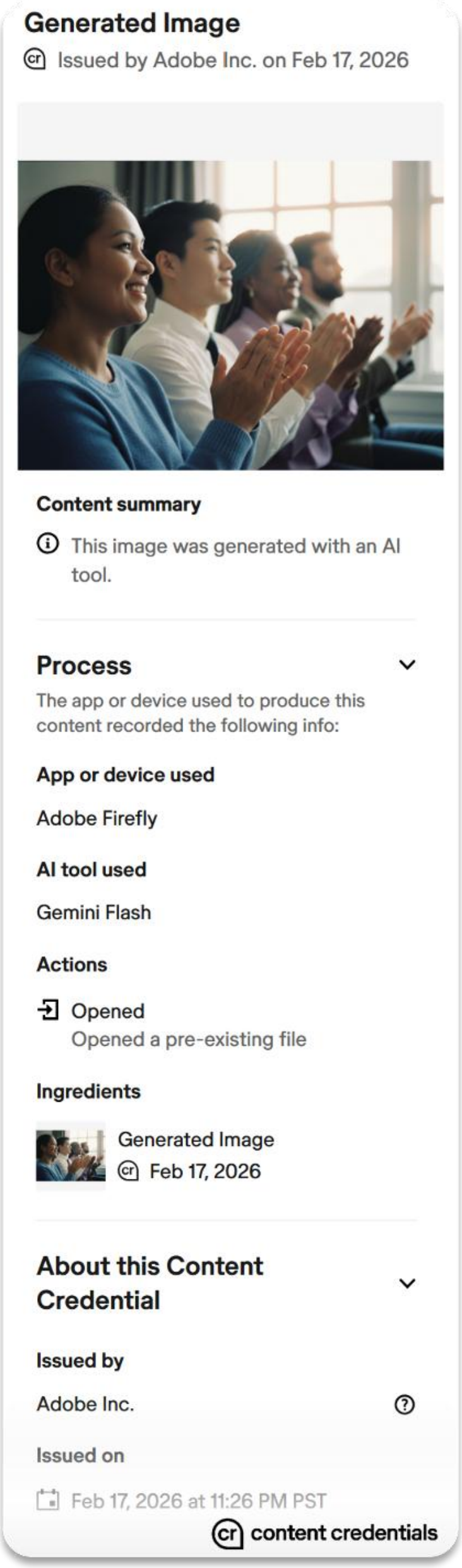

*Source: C2PA Verify Tool; Adobe Firefly (image)*

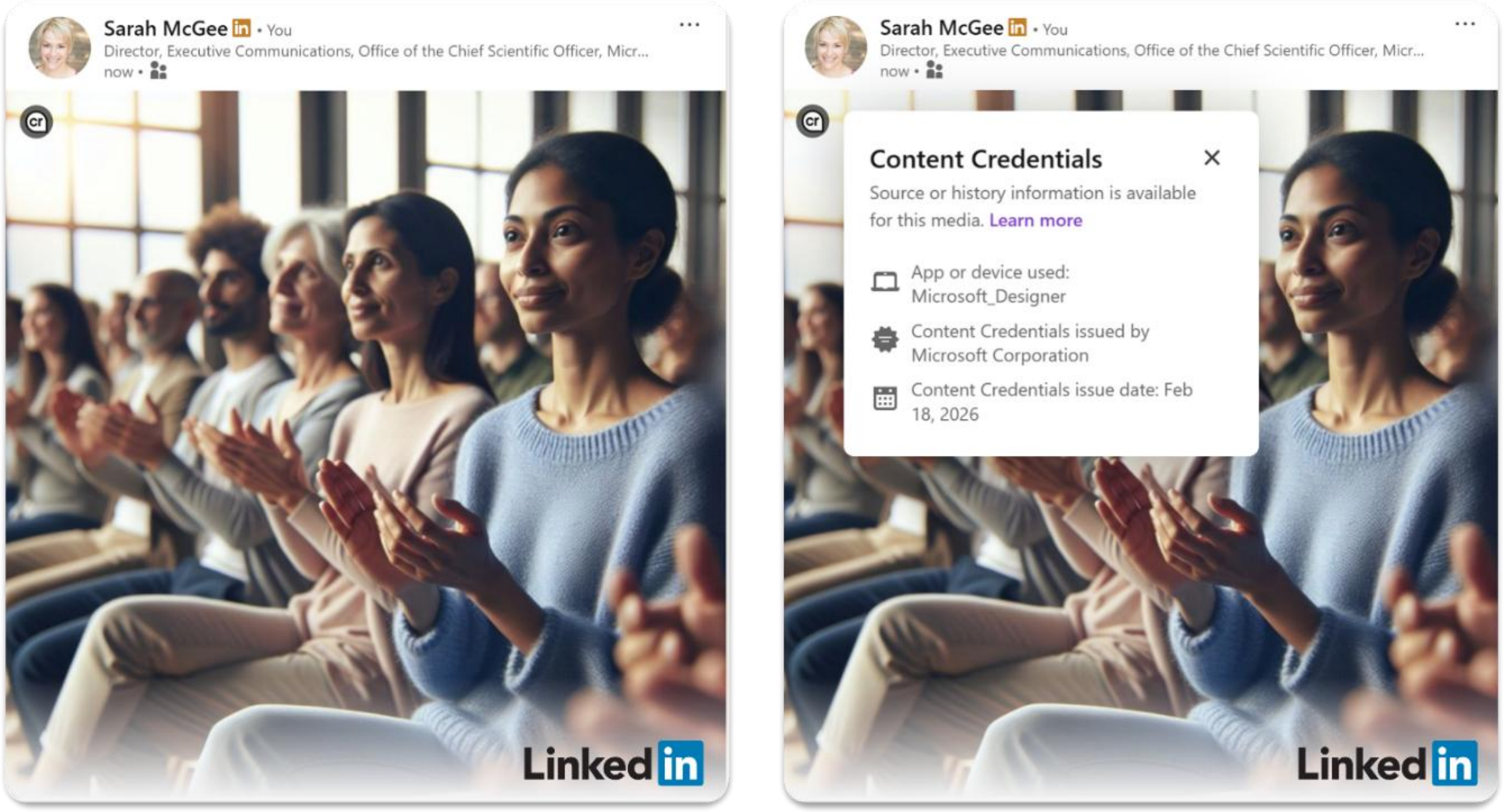

*Source: LinkedIn (L-R: Cr automatically displayed → select Cr icon to view provenance signals)*

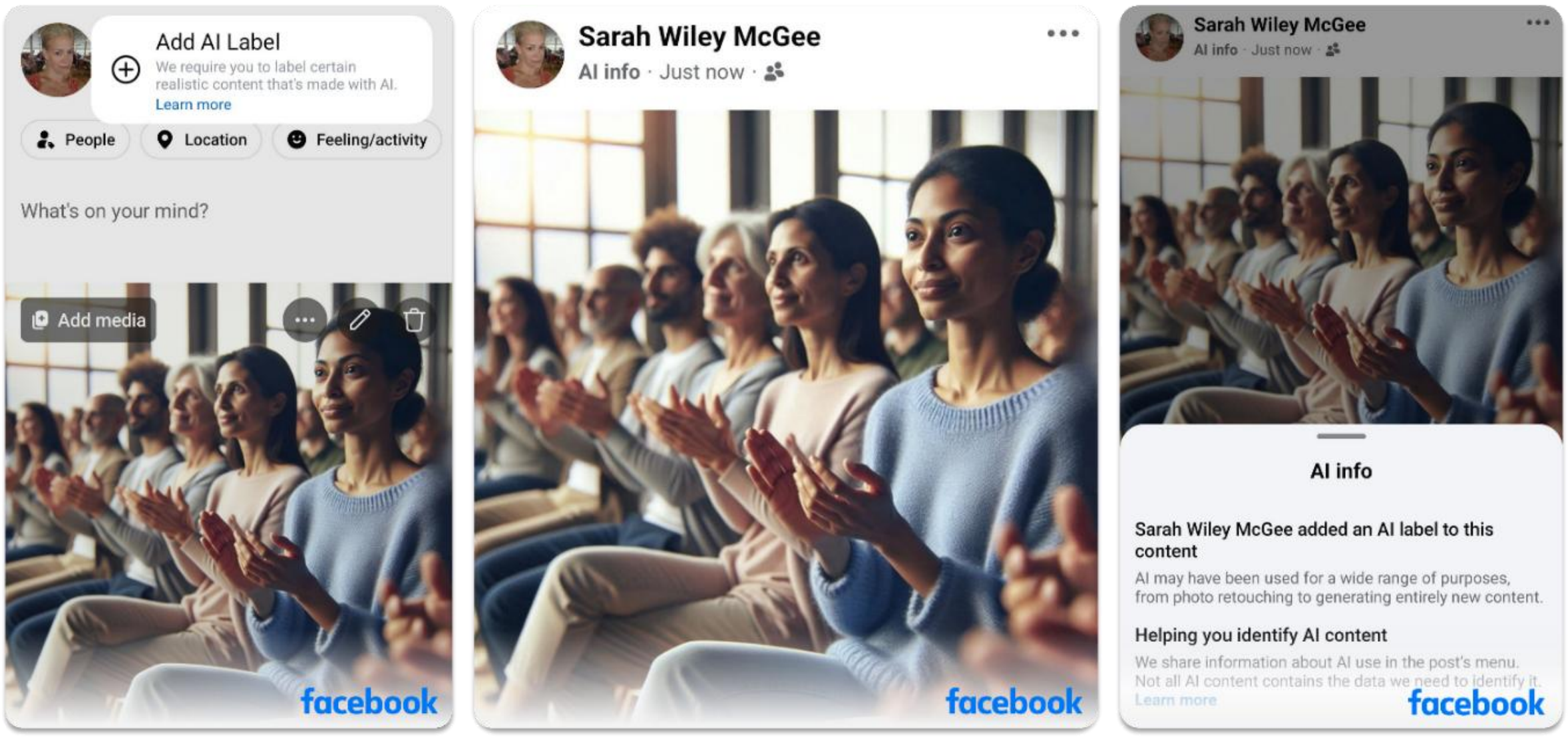

*Source: Facebook (L-R: user opts into AI label → AI info label appears → select AI info for additional information)*

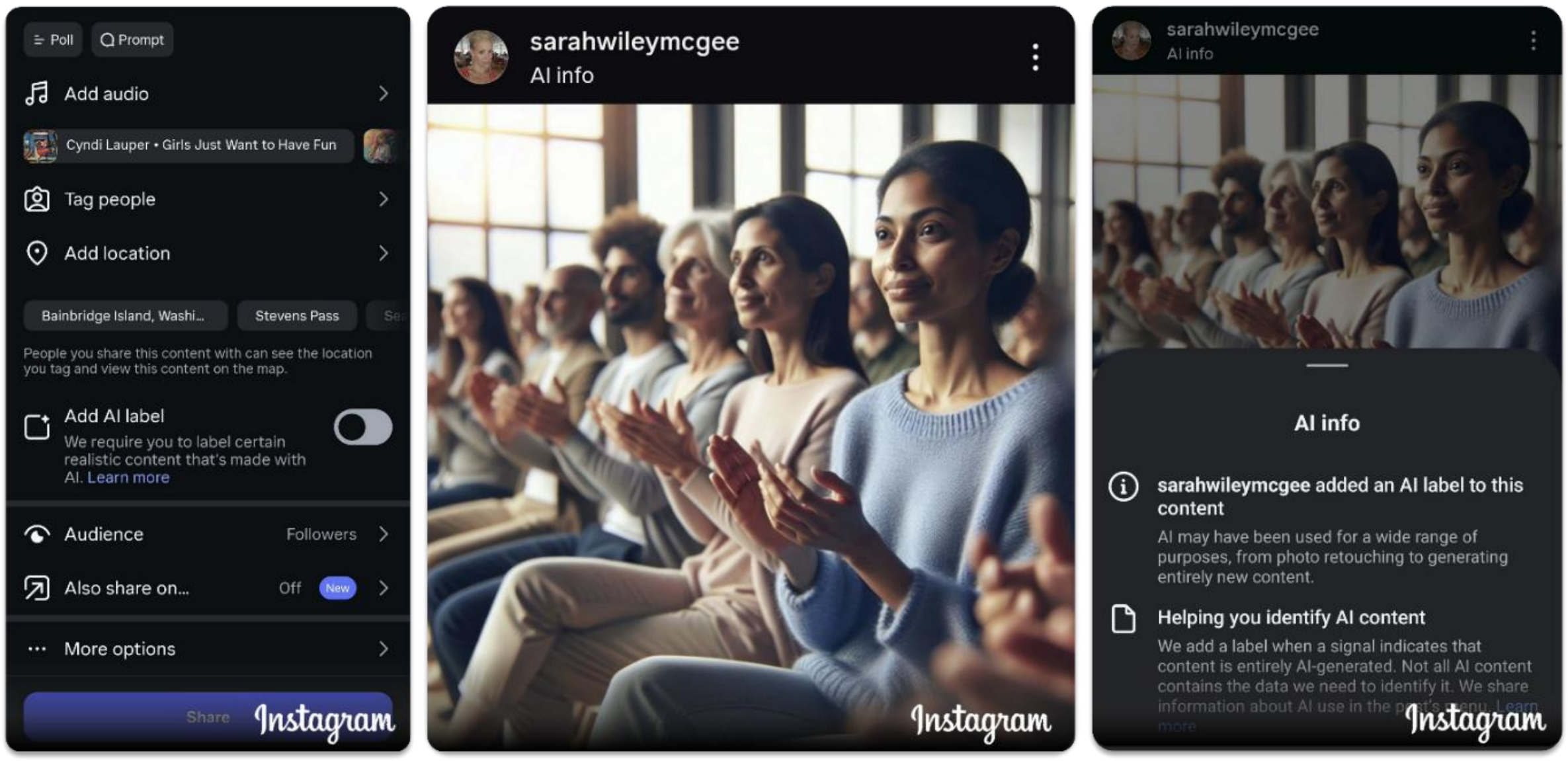

*Source: Instagram (L-R: user opts into AI label → AI info label appears → select AI info for additional information)*

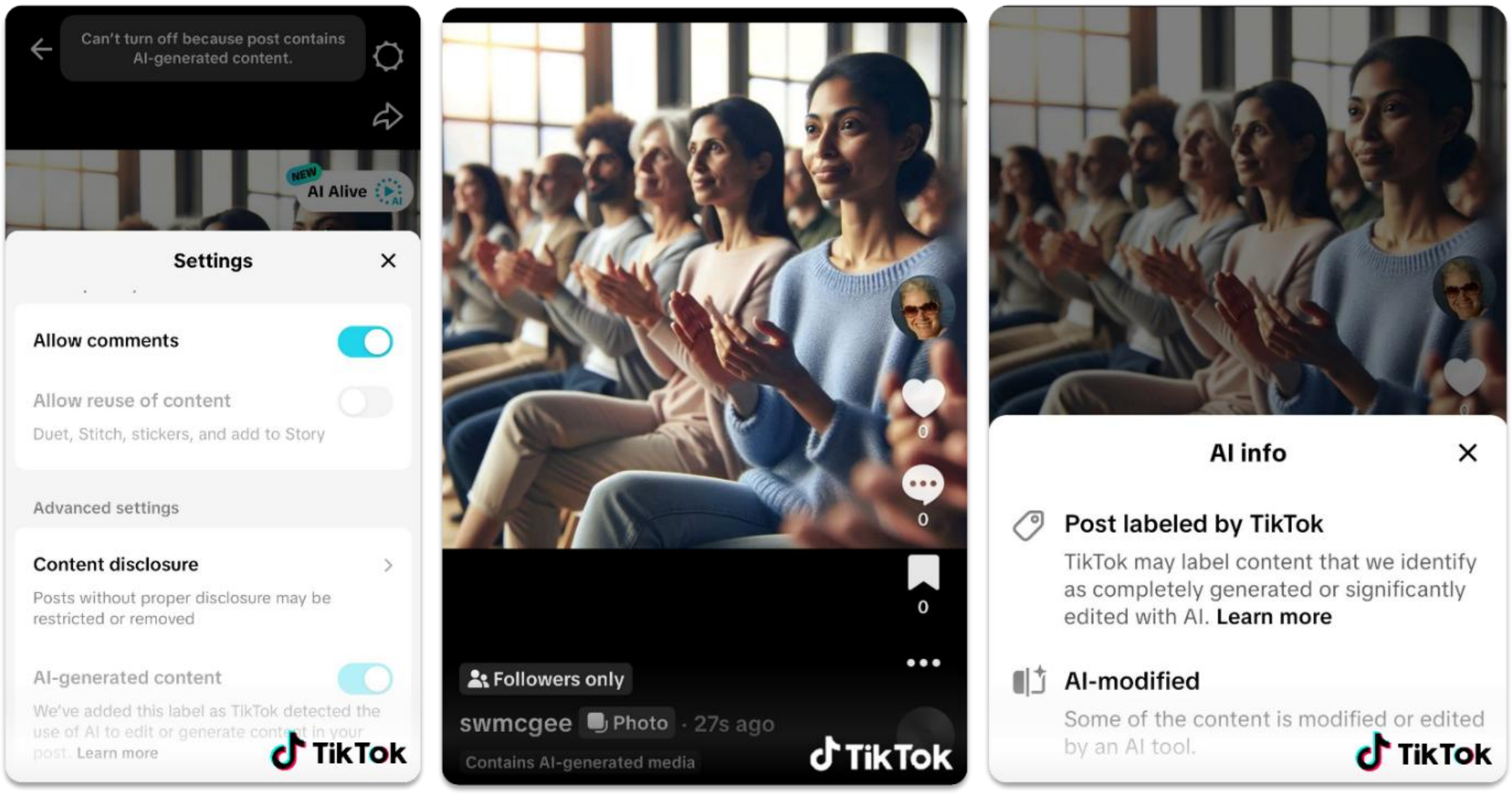

*Source: TikTok (L-R: user opts into disclosures → AI label appears → select AI content for more information)*

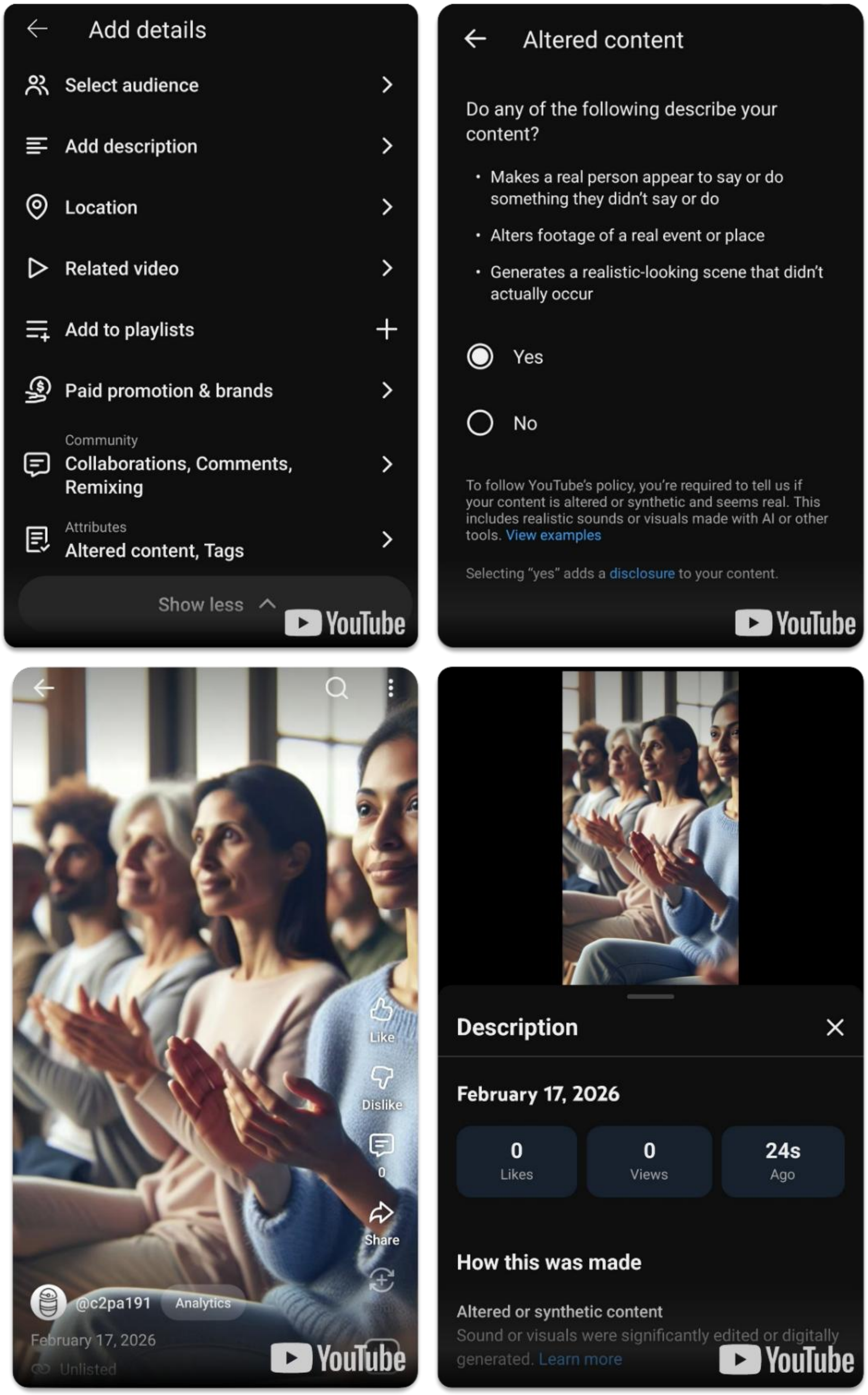

*Source: YouTube (L-R: user opts into AI label → user identifies tag → select ampersand to see more details).*